%% file: paper.tex
\documentclass[a4,12pt]{article}

\usepackage{amsmath,amssymb}

\makeatletter

\@addtoreset{equation}{section}
\makeatother
  
\def\tfontsize{scaled\magstep4}
\font\titlerm=cmr10 \tfontsize 

\def\CD{{\cal D}}
\def\CJ{{\cal J}}

\def\CN{{\cal N}}
\def\CO{{\cal O}}

\def\BQ{{\overline Q}}
\def\BD{{\overline D}}
\def\Bb{{\overline b}}

\def\BG{{\overline G}}

\def\BJ{{\overline J}}

\def\BU{{\overline U}}
\def\Bphi{{\overline \phi}}
\def\Bpsi{{\overline \psi}}
\def\Bxi{{\overline \xi}}

\def\Blambda{{\overline \lambda}}
\def\Bsigma{{\overline \sigma}}

\def\BPhi{{\overline \Phi}}

\def\BXi{{\overline \Xi}}
\def\Btheta{{\overline \theta}}
\def\BLambda{{\overline \Lambda}}
\def\BSigma{{\overline \Sigma}}
\def\BUpsilon{{\overline \Upsilon}}
\def\BOmega{{\overline \Omega}}

\def\BCD{{\overline \CD}}

\def\lnp{:\!}
\def\rnp{\!:}

\newcommand{\bR}{\mathbb{R}}

\begin{document}
%%%%%%%%%%
\begin{titlepage}
\begin{flushright}
KUNS-2018 \\
YITP-06-19 \\
{\tt hep-th/0605021}
\end{flushright}
\setlength{\baselineskip}{19pt}
\bigskip\bigskip\bigskip

\vbox{\centerline{\titlerm Superconformal Symmetry}
\bigskip
\centerline{\titlerm in Linear Sigma Model on Supermanifolds}}

\bigskip\bigskip\bigskip

\centerline{Shigenori Seki\footnote{\tt seki@gauge.scphys.kyoto-u.ac.jp}, 
	Katsuyuki Sugiyama\footnote{\tt sugiyama@phys.h.kyoto-u.ac.jp}
	and Tatsuya Tokunaga\footnote{\tt tokunaga@yukawa.kyoto-u.ac.jp}}

\bigskip\bigskip

{\it
\centerline{$^*$Department of Physics, Graduate School of Science}
\centerline{Kyoto University, Kyoto 606-8502, Japan}
\bigskip
\centerline{$^\dagger$Department of Physics, Graduate School of Science, Yoshida South}
\centerline{Kyoto University, Kyoto 606-8501, Japan}
\bigskip
\centerline{$^\ddagger$Yukawa Institute for Theoretical Physics}
\centerline{Kyoto University, Kyoto 606-8502, Japan}
}

\vskip .3in

\centerline{\bf abstract}

We consider a gauged linear sigma model in two dimensions with 
Grassmann odd chiral superfields. 
We investigate the Konishi anomaly of this model and find out 
the condition for realization of superconformal symmetry on 
the world-sheet.  When this condition is satisfied, 
the theory is expected  to flow into conformal theory in the infrared limit. 
We construct superconformal currents explicitly and 
study some properties of this world-sheet theory 
from the point of view of conformal field theories.

\vfill\noindent
May 2006
\end{titlepage}

%%%%%%%%%%
\setlength{\baselineskip}{19pt}

\section{Introduction}

It has recently proposed that the topological string theory on a 
supermanifold $\mathbb{CP}^{3|4}$ describes a 
twistor string \cite{Witten:2003nn}, which is associated 
with super Yang-Mills theory and 
conformal supergravities in four dimensions. 
This topological B-model provides a powerful prescription for computing 
amplitudes of the super Yang-Mills theory. 
 
This surprising correspondence attracts a lot of attention 
and various works have been done in this topic. 
Super Yang-Mills amplitudes are further 
investigated and interesting 
relations between the Yang-Mills theory and the topological string 
have been developed. On the other hand, string theories on 
the twistor spaces \cite{Gukov:2004ei}-\cite{Wolf:2004hp}, 
%\cite{Gukov:2004ei, Popov:2004rb, Lechtenfeld:2004cc, Popov:2004nk, Siegel:2004dj,
%  Sinkovics:2004fm, Movshev:2004ub, Wolf:2004hp}
\cite{Seki:2005hx, Tokunaga:2005pj} 
or supermanifolds themselves were studied 
to formulate mirror symmetry for supermanifolds \cite{Sethi:1994ch}-\cite{Kaura:2006hb} 
%\cite{Sethi:1994ch, Schwarz:1995ak, Aganagic:2004yh, Kumar:2004dj, Ahn:2004xs,
%  Belhaj:2004ts, AhlLaamara:2006rk, Kaura:2006hb} 
and related string dualities \cite{Berkovits:2004hg}-\cite{Lechtenfeld:2005xi},
%\cite{Berkovits:2004hg, Berkovits:2004tx, Berkovits:2004jj,
% Ahn:2004yu, Park:2004bw, Giombi:2004xv, Ahn:2004ua, Grassi:2004tv, Saemann:2005ji, Lechtenfeld:2005xi}, 
\cite{Neitzke:2004pf, Nekrasov:2004js}. 

It is interesting to study geometry of supermanifolds \cite{Rocek:2004bi, Zhou:2004su} and 
to find profound structures beyond the bosonic manifolds. 
We investigated critical string theories on 
flat supermanifolds \cite{Tokunaga:2005pj}
toward construction of consistent strings on these manifolds. 
It was a first step to understand the 
strings on the superspace as world-sheet theories and 
to obtain consistency conditions for them. 
Also we analyzed supermanifolds 
from the point of view of the 
gauged linear sigma models \cite{Witten:1993yc}, \cite{Sethi:1994ch, Schwarz:1995ak, Aganagic:2004yh, Seki:2005hx} 
in order to understand properties of curved superspace backgrounds. 
For backgrounds for usual bosonic manifolds, 
Ricci-flat condition is necessary to 
construct consistent vacua of strings . 
If this condition is satisfied, 
conformal theories are realized on the world-sheet 
and physics of infrared (IR) region is controlled by 
conformal algebra. 
For models with supersymmetry, 
superconformal currents play important roles in 
describing physical properties of the IR dynamics.

In this paper, we shall consider gauged linear sigma models  
on weighted projective superspaces and 
investigate geometrical properties 
from viewpoints of world-sheet theories. 

In section~2, we review gauged linear sigma models with a $U(1)$ gauge field 
and chiral superfields shortly. 
In section~3, we explain $U(1)$ R-symmetries and 
conservation of associated currents at the classical level. 
But one of the $U(1)$ symmetries is broken at the quantum level and 
induces anomaly. In section~4, we investigate this Konishi anomaly \cite{Konishi:1983hf,Konishi:1985tu}
for this $U(1)$ symmetry and construct anomaly equations explicitly. 
In deriving this result, we take two approaches; 
covariant calculation by the path integral methods and 
evaluation of one-loop effects in the light-cone gauge. 
Then we obtain the condition of anomaly cancellation for 
this symmetry and construct conserved currents explicitly. 
These supersymmetric currents  flow into 
superconformal currents in the IR limit if 
there is no anomaly. 
In section~5, we calculate operator products of these currents and 
study superconformal algebra of the IR theory. 
The lowest component of the field strength of 
vector superfield plays essential roles in these 
curved backgrounds. 
Together with gaugino, they behave as ghost fields. 
Also we discuss holomorphic forms and associated  
extended algebra for a few concrete cases.
Section~6 is devoted to conclusions and discussions.

\section{Model}

In this section we review $\CN =2$ gauged linear sigma model 
shortly \cite{Witten:1993yc}.
We want to describe supermanifolds by using 
two dimensional gauged linear sigma models.
For simplicity we consider a $U(1)$ gauge group. 
The supercoordinates of the world-sheet are 
denoted by $(x_0,x_1,\theta^\pm,\Btheta^\pm)$. We set 
the metric of the world-sheet to be $\eta_{ij} = {\rm diag}(-1,+1)$.
This model has $\CN =2$ supersymmetry on the world-sheet and  
there are four types of supercharges $Q_{\pm}$ and $\BQ_{\pm}$
\begin{equation}
Q_\pm = {\partial \over \partial\theta^\pm} 
+ i\Btheta^\pm(\partial_0\pm\partial_1), \quad
\BQ_\pm = -{\partial \over \partial\Btheta^\pm} 
- i\theta^\pm(\partial_0\pm\partial_1) . 
\end{equation}
Associated superderivatives $D_{\pm}$ and $\BD_{\pm}$ are defined:
\begin{equation}
D_\pm = {\partial \over \partial\theta^\pm} - i\Btheta^\pm(\partial_0\pm\partial_1), \quad
\BD_\pm = -{\partial \over \partial\Btheta^\pm} + i\theta^\pm(\partial_0\pm\partial_1) . 
\end{equation}

We shall take 
$m$ bosonic and $n$ fermionic coordinates and consider 
an $(m|n)$-dimensional superspace.
Since these coordinates are regarded as the lowest components 
of chiral superfields, we introduce $m$ bosonic chiral superfields and 
$n$ fermionic chiral superfields,
\[
\Phi^I_0(x),\quad (I=1,2,\cdots,m),\qquad \Xi^A_0(x),\quad (A=1,2,\cdots,n).
\]

Firstly an ordinary bosonic chiral superfield $\Phi^I_0$  
is defined by $\BD \Phi^I_0 = 0$ and expressed in terms of component fields
as 
\begin{eqnarray}
\Phi^I_0 &=& \phi^I + \sqrt{2}(\theta^+\psi^I_+ + \theta^-\psi^I_-)
+ 2\theta^+\theta^- F^I \nonumber \\
&&-i\theta^-\Btheta^-(\partial_0-\partial_1)\phi^I 
-i\theta^+\Btheta^+(\partial_0+\partial_1)\phi^I 
- \theta^+\theta^-\Btheta^-\Btheta^+ (\partial_0^2 - \partial_1^2)\phi^I \nonumber \\
&&-i\sqrt{2}\theta^+\theta^-\Btheta^-(\partial_0-\partial_1)\psi^I_+
+i\sqrt{2}\theta^+\theta^-\Btheta^+(\partial_0+\partial_1)\psi^I_- .
\end{eqnarray}
Here $\phi^I$ and $F^I$ are bosons, while $\psi^I_+$ and $\psi^I_-$ are fermions. 
In the same way, we can define the fermionic chiral superfield $\Xi^A_0$
in component expansion: 
\begin{eqnarray}
\Xi^A_0 &=& \xi^A + \sqrt{2}(\theta^+ b^A_+ + \theta^- b^A_-)
+ 2\theta^+\theta^- \chi^A \nonumber \\
&&-i\theta^-\Btheta^-(\partial_0-\partial_1)\xi^A 
-i\theta^+\Btheta^+(\partial_0+\partial_1)\xi^A 
- \theta^+\theta^-\Btheta^-\Btheta^+ (\partial_0^2 - \partial_1^2)\xi^A \nonumber \\
&&-i\sqrt{2}\theta^+\theta^-\Btheta^-(\partial_0-\partial_1)b^A_+
+i\sqrt{2}\theta^+\theta^-\Btheta^+(\partial_0+\partial_1)b^A_- .
\end{eqnarray}
Fields $\xi^A$ and $\chi^A$ are Grassmann odd, 
while $b^A_+$ and $b^A_-$ are Grassmann even. 
So this chiral superfield behaves totally as a Grassmann odd superfield.

In order to construct a projective supermanifold, 
we consider the $U(1)$ gauge theory 
by introducing an abelian vector superfield $V$.
It is written in the Wess-Zumino gauge,
\begin{eqnarray}
V &=& -\sqrt{2}\theta^-\Btheta^+\sigma - \sqrt{2}\theta^+\Btheta^-\Bsigma
+ \theta^-\Btheta^-(v_0-v_1) + \theta^+\Btheta^+(v_0+v_1) \nonumber \\
&&- 2i\theta^+\theta^-(\Btheta^+\Blambda_+ + \Btheta^-\Blambda_-)
-2i\Btheta^-\Btheta^+(\theta^+\lambda_++\theta^-\lambda_-)
+ 2\theta^+\theta^-\Btheta^-\Btheta^+D , 
\end{eqnarray}
and associated field strength of the superfield  
$\Sigma = (1 / \sqrt{2})\BD_+D_-V$ is represented as
\begin{eqnarray*}
\Sigma &=& \sigma + i\sqrt{2}\theta^+\Blambda_+ - i\sqrt{2}\Btheta^-\lambda_-
+ \sqrt{2}\theta^+\Btheta^- (D - iv_{01}) 
+ i\theta^-\Btheta^-(\partial_0-\partial_1)\sigma \\
&&-i\theta^+\Btheta^+(\partial_0+\partial_1)\sigma 
- \sqrt{2}\theta^+\theta^-\Btheta^-(\partial_0-\partial_1)\Blambda_+
+ \sqrt{2}\theta^+\Btheta^-\Btheta^+(\partial_0+\partial_1)\lambda_- \\
&&+\theta^+\theta^-\Btheta^-\Btheta^+ (\partial_0^2-\partial_1^2)\sigma ,\\
&&v_{01} \equiv \partial_0v_1 - \partial_1v_0. 
\end{eqnarray*}
This field strength is the twisted chiral superfield and 
the kinetic part of the gauge field is denoted by
\[
L_{\rm gauge} = - {1 \over e^2}\int d^4\theta\ \BSigma\Sigma 
\]
where $e$ is a gauge coupling constant.
We then assign $U(1)$ charges $Q_I$ on $\Phi^I_0$ and $q_A$ on $\Xi^A_0$.
Then ungauged $(m|n)$-dimensional target space 
is reduced to a projective manifold 
due to this $U(1)$ gauge field.
Now let us write down the Lagrangian of 
${\cal N} = (2,2)$ gauged linear sigma model
\begin{eqnarray*}
&&L = L_{\rm kin} + L_{\rm gauge} , \\
&&L_{\rm kin} = \int d^4\theta \left( \sum_{I=1}^m\BPhi^I_0e^{2Q_IV}\Phi^I_0+
\sum_{A=1}^n\BXi^A_0e^{2q_AV}\Xi^A_0\right).
\end{eqnarray*}
When one considers the IR limit, 
this model turns to describe physics of the sigma model with target space 
$\mathbb{WCP}^{m-1|n}_{(Q_1,Q_2,\cdots ,Q_m|q_1,q_2,\cdots ,q_n)}$,  
namely $(m-1|n)$-dimensional projective space.

\section{Current and $U(1)$ Symmetry}

The physics of the model in the IR region depends on the 
charges $(Q_I,q_A)$. When $\sum_I Q_I -\sum_A q_A>0$, 
the model is asymptotic free. 
If $\sum_I Q_I -\sum_A q_A =0$ is satisfied, the 
theory is scale invariant in the one-loop approximation and 
we can expect to construct conformal currents of the 
underlying IR theory. 
We shall discuss this conformal theory.
Now we introduce a current $\CJ$, 
\begin{eqnarray}
\CJ &=& {1 \over 2}\sum_I D_- \Bigl(\Phi^I_0 e^{2Q_I V}\Bigr) e^{-2Q_I V} \BD_- \Bigl(e^{2Q_I V} \BPhi^I_0\Bigr) \nonumber \\
&&+ {1 \over 2}\sum_A D_- \Bigl(\Xi^A_0 e^{2q_A V}\Bigr) e^{-2q_A V} \BD_- 
\Bigl(e^{2q_A V} \BXi^A_0\Bigr) 
+ {2i \over e^2} \Sigma \partial_-\BSigma . \label{Eq:current}
\end{eqnarray}
We can verify conservation of the current classically $\BD_+ \CJ = 0$ 
by using equations of motion, 
\begin{eqnarray}
\BD_+\BD_- \Bigl(\BPhi^I_0e^{2Q_I V}\Bigr) &=& 0 , \\
\BD_+\BD_-\Bigl(\BXi^A_0e^{2q_A V}\Bigr) &=& 0 , \\
\sum_I Q_I \BPhi^I_0 e^{2Q_I V} \Phi^I_0 
	+\sum_A q_A \BXi^A_0 e^{2q_A V} \Xi^A_0 
	&=& {1 \over 2\sqrt{2}e^2}\bigl(\BD_-D_+\Sigma + \BD_+D_-\BSigma\bigr).
\end{eqnarray}
This current (\ref{Eq:current}) has expansion in terms of supercoordinates 
$\theta^{\pm}$, $\Btheta^{\pm}$.
In order to write down formulae of currents explicitly, we 
expand superfields in terms of $\theta^-$ and $\Btheta^-$
\begin{eqnarray*}
\Phi^I_0 &=& {\Phi'}^I_0+\sqrt{2}\theta^-\Lambda^I_{-0}-2i\theta^-\Btheta^-\partial_-{\Phi'}^I_0 , \\
\Xi^A_0 &=& {\Xi'}^A_0+\sqrt{2}\theta^-{\tilde \Lambda}^A_{-0}
	-2i\theta^-\Btheta^-\partial_-{\Xi'}^A_0 , \\
V &=& \Psi+\theta^-\Btheta^-(v_0-v_1-2ia)-\sqrt{2}\theta^-\Btheta^+\Sigma'
	-\sqrt{2}\theta^+\Btheta^-\BSigma' ,\\
a &=& \theta^+\Blambda_-+\Btheta^+\lambda_-+i\theta^+\Btheta^+D , \\
\Sigma &=& \Sigma'+{i\over \sqrt{2}}\Btheta^-\Upsilon 
	+2i\theta^-\Btheta^-\partial_-\Sigma' , \\
&&\partial_{\pm} \equiv \frac{1}{2}(\partial_0\pm \partial_1) ,
\end{eqnarray*}
where component fields are defined as
\begin{eqnarray*}
{\Phi'}^I_0 &=& \phi^I +\sqrt{2}\theta^+\psi^I_+
	-2i\theta^+\Btheta^+\partial_+\phi^I , \\
{\Xi'}^A_0 &=& \xi^A +\sqrt{2}\theta^+ b^A_+
	-2i\theta^+\Btheta^+\partial_+\xi^A , \\
\Lambda^I_{-0} &=& \psi^I_--\sqrt{2}\theta^+F^I
	-2i\theta^+\Btheta^+\partial_+\psi^I_- , \\
{\tilde \Lambda}^A_{-0} &=& b^A_--\sqrt{2}\theta^+\chi^A 
	-2i\theta^+\Btheta^+\partial_+ b^A_- , \\
\Psi &=& \theta^+\Btheta^+ (v_0 - v_1), \\
\Sigma' &=& \sigma +i\sqrt{2}\theta^+\Blambda_+
	-2i\theta^+\Btheta^+\partial_+\sigma , \\
\Upsilon &=& -2\lambda_-+2i\theta^+(D-iv_{01})
	+4i\theta^+\Btheta^+\partial_+\lambda_- .
\end{eqnarray*}
If we introduce fields defined as follows: 
\begin{eqnarray*}
{\Phi'}^I &=& e^{Q_I\Psi}{\Phi'}^I_0\,,\quad 
{\Xi'}^A=e^{q_A\Psi}{\Xi'}^A_0 ,\\
\Lambda^I_- &=& e^{Q_I\Psi}(\Lambda^I_{-0}-2Q_I\Btheta^+\Sigma'{\Phi'}^I_0) , \\
\tilde{\Lambda}^A_- &=& e^{q_A\Psi}(\tilde{\Lambda}^A_{-0}-2q_A\Btheta^+\Sigma'{\Xi'}^A_0) , \\ 
(\CD_0-\CD_1){\Phi'}^I &=& (\partial_0-\partial_1+iQ_I(v_0-v_1-2ia)){\Phi'}^I , \\
(\CD_0-\CD_1){\Xi'}^A &=& (\partial_0-\partial_1+iq_A(v_0-v_1-2ia)){\Xi'}^I ,
\end{eqnarray*}
then component expansion of $\CJ$ is represented 
by currents $(J,G,\BG,T)$
\begin{eqnarray}
\CJ &=& J+2\sqrt{2}i\theta^-G+2\sqrt{2}i\Btheta^-\BG +4\theta^-\Btheta^-T ,\nonumber \\
J &=& \sum_I \Lambda^I_-\BLambda^I_- 
	-\sum_A \tilde \Lambda^A_-\overline{{\tilde \Lambda}}^A_-
	+{2i\over e^2}\Sigma'\partial_-\BSigma' ,\label{currentj}\\
G &=& -\sum_I\frac{1}{2}\Lambda^I_-(\BCD_0-\BCD_1)\BPhi'^I
	+\sum_A\frac{1}{2}{\tilde \Lambda}^A_-(\BCD_0-\BCD_1)\BXi'^A
	+{i\over 2e^2}\Sigma' \partial_-\BUpsilon ,\label{currentg}\\
\BG &=& -\sum_I\frac{1}{2}(\CD_0-\CD_1)\Phi'^I\cdot \BLambda_-^I
	+\sum_A\frac{1}{2}(\CD_0-\CD_1)\Xi'^A \cdot \overline{\tilde \Lambda}_-^A
	+{i\over 2e^2}\Upsilon\partial_-\BSigma' , \label{currentbg}\\
T &=& -\sum_I\frac{1}{2}(\CD_0-\CD_1)\Phi'^I(\BCD_0-\BCD_1)\BPhi'^I 
	+\sum_A\frac{1}{2}(\CD_0-\CD_1)\Xi'^A(\BCD_0-\BCD_1)\BXi'^A \nonumber \\
	&&-{i\over 4e^2}\Upsilon\partial_-\BUpsilon 
	-{1\over e^2}\partial_-\Sigma'\partial_-\BSigma'
	+{1\over e^2}\Sigma'\partial_-^2\BSigma' \nonumber \\
	&&-{i\over 4}\sum_I \Bigl[\Lambda^I_-(\BCD_0-\BCD_1)\BLambda^I_- 
	-(\CD_0-\CD_1)\Lambda^I_-\cdot \BLambda^I_- \Bigr]\nonumber \\
	&&+{i\over 4}\sum_A\Bigl[{\tilde \Lambda}^A_-(\BCD_0-\BCD_1)\overline{\tilde \Lambda}^A_- 
	-(\CD_0-\CD_1){\tilde \Lambda}^A_- \cdot \overline{\tilde \Lambda }^A_- \Bigr] .\label{currentt}
\end{eqnarray}
These are grouped into an  $\CN=2$ multiplet 
and the current in the highest component is 
identified with the energy momentum tensor. 
If we concentrate on the left movers, 
then component fields are expressed explicitly
\begin{eqnarray}
J &=& \sum_I \psi_-^I \Bpsi_-^I - \sum_A b_-^A \Bb_-^A
		+{2i \over e^2}\sigma \partial_-\Bsigma , \label{Eq:J1st}\\
G &=& -\sum_I \psi_-^I \BCD_- \Bphi^I 
		+\sum_A b_-^A \BCD_- \Bxi^A 
		-{i \over e^2}\sigma \partial_-\Blambda_- , \\
\BG &=& -\sum_I \CD_- \phi^I \cdot \Bpsi_-^I  
		+\sum_A \CD_- \xi^A \cdot \Bb_-^A 
		-{i \over e^2}\lambda_- \partial_-\Bsigma , \\
T &=& -\sum_I 2 \BCD_- \Bphi^I \CD_- \phi^I 
		+{i \over 2}\sum_I(\CD_- \psi_-^I \cdot \Bpsi_-^I 
		-\psi_-^I\BCD_- \Bpsi_-^I )  \nonumber \\
	&&+\sum_A 2 \BCD_- \Bxi^A \CD_- \xi^A 
		-{i\over 2}\sum_A(\CD_- b_-^A \cdot \Bb_-^A 
		-b_-^A \BCD_-\Bb_-^A) \nonumber \\
	&&-{i \over e^2}\lambda_- \partial_-\Blambda_- 
		-{i\over e^2}(\partial_-\sigma\partial_-\Bsigma
                -\sigma\partial_-^2\Bsigma) , \label{Eq:J4th}
\end{eqnarray}
where $\CD_- \equiv (1/2)(\CD_0 - \CD_1)|_{a=0}$.
They could be superconformal currents if there is no anomaly.

Our model has two $U(1)$ R-symmetries $U(1)_R$ and $U(1)_L$ at the classical level.
They act on the supercoordinates 
$(\theta^+,\theta^-,\Btheta^+,\Btheta^-)$ 
\[
\begin{array}{l}
U(1)_R\,;\,(e^{+i\alpha}\theta^+,\theta^-,e^{-i\alpha}\Btheta^+,\Btheta^-)\,,\\
U(1)_L\,;\,(\theta^+,e^{+i\beta}{\theta}^-,\Btheta^+,e^{-i\beta}\Btheta^-)\,,\quad
\alpha\,,\beta\in {\bR}\,.
\end{array}
\]
But they are generally anomalous at the quantum level. 
In the next section, we shall discuss this anomaly.

\section{Anomaly}

We shall look at the $U(1)_L$ symmetry.
It acts on coordinates $(\theta^-,\Btheta^-)$ as 
$(e^{+i\beta}\theta^-,e^{-i\beta}\Btheta^-)$. 
Then relevant superfields $(\Lambda_-,\tilde{\Lambda}_-,\Sigma')$ transform 
to $e^{-i\beta}(\Lambda_-,\tilde{\Lambda}_-,\Sigma')$. 
This $U(1)_L$ symmetry is conserved classically, 
but broken at quantum level 
and induces the Konishi anomaly \cite{Konishi:1983hf,Konishi:1985tu}.
Now we compute this Konishi anomaly.  

First we can rewrite Lagrangian by using component expansion
\begin{eqnarray*}
S &=& \frac{1}{2}\int d^2y \int d\theta^+d\Btheta^+ 
\biggl[
\sum_I\BLambda^I_-\Lambda^I_-
+\sum_A\overline{\tilde{\Lambda}}^A_-\tilde{\Lambda}^A_-
+\frac{i}{e^2}(\BSigma'\partial_-\Sigma'-\partial_-\BSigma'\cdot\Sigma') \\
&&+\sum_I\frac{i}{2}\BPhi'^I({\cal D}_0-{\cal D}_1){\Phi'}^I
-\sum_I\frac{i}{2}(\BCD_0-\BCD_1)\BPhi'^I\cdot {\Phi'}^I \\
&&+\sum_A\frac{i}{2}\BXi'^A({\cal D}_0-{\cal D}_1){\Xi'}^A
-\sum_A\frac{i}{2}(\BCD_0-\BCD_1)\BXi'^A\cdot {\Xi'}^A
+\frac{1}{4e^2}\BUpsilon \Upsilon
\biggr] . 
\end{eqnarray*}
In order to evaluate the anomaly, 
we consider transformations of the relevant fields under 
chiral $U(1)$ symmetry\footnote{For simplicity, 
we abbreviate flavor indices ``$I$'' and ``$A$'' for a moment} 
\begin{equation}
( \Lambda_- , {\tilde{\Lambda}}_-  , {\Sigma}' ) \to e^{i A} 
( \Lambda_- , {\tilde{\Lambda}}_-  , {\Sigma}' ) \equiv ({\hat{\Lambda}}_- , 
{\hat{{\tilde{\Lambda}}}}_- , {\hat{{\Sigma}}}' ) .
\end{equation}
Here the chiral superfield $A$ satisfies the 
condition $\BD_+ A =0$.  
The infinitesimal transformation $\delta S$ of the action 
can be expressed as 
\begin{equation}
\delta S = \frac{1}{2} \int d^2 y \int d \theta^+ 
	A ( - i \BD_+  ) \biggl[ \BLambda_- \Lambda_- 
	+{\overline{\tilde{\Lambda}}}_- {\tilde{\Lambda}}_- 
	-\frac{2i}{e^2} (\partial_- {\BSigma}') {\Sigma}' \biggr] 
\equiv -\frac{1}{2} \int d^2 y \int d \theta^+
A(-i\BD_+ )J . 
\end{equation}
Then the partition function $Z$ is represented under this $U(1)$ symmetry
\begin{eqnarray}
Z &=& \int D [\Lambda_- ,{\tilde{\Lambda}}_- ,{\Sigma}']\, e^{i S}  \nonumber \\
&=& \int D [\Lambda_- ,{\tilde{\Lambda}}_- ,{\Sigma}']
\,{\rm Sdet} \frac{({\hat{\Lambda}}_- ,{\hat{{\tilde{\Lambda}}}}_- ,{\hat{\Sigma}}')}  
{(\Lambda_- , {\tilde{\Lambda}}_- , {\Sigma}')}\,e^{i(S +\delta S)} .
\end{eqnarray}
In this formula, $J$ is the same current 
as the one given in Eq.(\ref{currentj}).
Superdeterminant Sdet$(\cdots)$ is the Jacobian of this transformation
\begin{equation}
{\rm Sdet} \frac{ ({\hat{\Lambda}}_- , {\hat{{\tilde{\Lambda}}}}_- , {\hat{{\Sigma}}}' ) }  
{ (\Lambda_- , {\tilde{\Lambda}}_- , {\Sigma}')} = {\rm Sdet} ( - i  A \BD_+ ) 
= e^{{\rm Str} ( - i A \BD_+ ) }. \label{Jac}
\end{equation}
If this Jacobian is not identity, it induces anomaly. 
We can evaluate this effect by the Fujikawa's method.
In doing this calculation actually, 
we need to regularize the superdeterminant and 
introduce the regulator $L$ \cite{Basu:2003bq};
\begin{eqnarray*}
L &=& - \frac{i}{2} \BD_+ e^{- Q\Psi} ( {\cal D}_0 - {\cal D}_1 ) 
 e^{- Q\Psi} \BD_+ e^{2 Q\Psi}  \\
&=& - \frac{i}{2} Q \Upsilon e^{-2 Q \Psi } D_+ e^{2 Q \Psi} 
 + e^{- Q \Psi } ( {\cal D}_0 - {\cal D}_1 ) ( {\cal D}_0 + {\cal D}_1 ) e^{Q \Psi}  \\
&&+ \frac{i}{2} e^{- Q \Psi } ( {\cal D}_0 - {\cal D}_1 )
 {\cal D}_+ \BCD_+ e^{Q \Psi} . 
\end{eqnarray*}
In deriving the formula in the second line,  we used a relation 
\begin{equation}
Q \Upsilon = \left[\, \BCD_+ ,  \BCD_0 - \BCD_1 \right] .
\end{equation}
Then we are able to evaluate 
the Jacobian (\ref{Jac}) by using this regulator
\begin{eqnarray}
e^{{\rm Str} ( - i A \BD_+ )} &=& \lim_{M \rightarrow \infty}
 \exp[ {\rm Str} ( - i A e^{\frac{L}{M^2} } \BD_+  ) ] , \nonumber \\
&\equiv& \lim_{M \rightarrow \infty} \exp \left[ \int d^2 y
 \int d \theta^+  \langle y, \theta^+, {\Btheta}^+ |
 ( - i A e^{\frac{L}{M^2} } \BD_+  ) | y, 
\theta^+, {\Btheta}^+ \rangle \right] .
\end{eqnarray}
First we compute the contribution from $\Lambda_-$.  
We insert a complete set in the integrand and evaluate the limit
\begin{eqnarray*}
&& \lim_{M \to \infty}  \langle y, \theta^+, {\Btheta}^+ |
 ( - i A e^{\frac{L}{M^2} } \BD_+  ) | y, \theta^+, {\Btheta}^+ \rangle \\
&=& \lim_{M \to \infty}  \int \frac{d^2 k}{(2 \pi)^2}   d \eta^+ d {\overline{\eta}}^+ 
\langle y, \theta^+, {\Btheta}^+ | ( - i A e^{\frac{L }{M^2}} \BD_+  ) | k,
 \eta^+, {\overline{\eta}}^+ \rangle 
\langle k, \eta^+, {\overline{\eta}}^+ | y, \theta^+, {\Btheta}^+ \rangle \\
&=& \lim_{M \to \infty}  \int \frac{d^2 k}{(2 \pi)^2}   d \eta^+ d {\overline{\eta}}^+
  e^{- i (k y + \eta \theta) } ( - i A e^{\frac{L}{M^2} } \BD_+  ) 
e^{i (k y + \eta \theta) } \\
&=& \lim_{M \to \infty}  \int \frac{d^2 k}{(2 \pi)^2} 
  d \eta^+ d {\overline{\eta}}^+  (- iA) ( i {\overline{\eta}}_+ - \theta^+ k_+ ) 
\, \exp \left[ - \frac{Q}{2 M^2} \Upsilon \eta_+  +\tilde{L} \right] \\
&=&  \frac{Q}{2 (2 \pi)^2} A \Upsilon \int d^2 \tilde{k} ~e^{- {\tilde{k}}_+ {\tilde{k}}_-  } 
= \frac{i}{8 \pi} QA \Upsilon . 
\end{eqnarray*}
Here we defined the rescaled momenta $(\tilde{k}_+,\tilde{k}_-)$ and $\tilde{L}$ as 
\begin{eqnarray*}
&& k_1 = M {\tilde{k}}_1 , \quad  k_2 = M {\tilde{k}}_2 , \\
&& \tilde{L} = - \frac{i}{2} \frac{Q}{M^2} \Upsilon (D_+ + 4 Q \partial_+ 
\Psi + {\Btheta}^+ k_+ ) \\
&& \qquad  + \frac{1}{M^2} ({\cal D}_0 - {\cal D}_1 + 2Q \partial_- \Psi + i k_-)
 ({\cal D}_0 + {\cal D}_1 + 2Q \partial_+ \Psi + i k_+)  .
\end{eqnarray*}
Finally, after summing up contributions from ${\tilde{\Lambda}}_-$ 
and recovering flavor indices, 
we obtain the Konishi anomaly equation 
\begin{equation}
\BD_+ J =  -\frac{1}{4 \pi } \biggl(\sum_IQ_I-\sum_Aq_A\biggr) \Upsilon .
\end{equation}
We can also rewrite the Konishi anomaly equation for the current ${\cal J}$ 
given by (\ref{Eq:current}),
\begin{equation}
\BD_+ {\cal J} = \frac{i \sqrt{2} }{4 \pi} \biggl(\sum_IQ_I -\sum_Aq_A \biggr)  \BD_- \Sigma .
\end{equation}
{}From this equation, 
we can see that the Konishi anomaly vanishes only if $\sum_IQ_I -\sum_Aq_A=0$.  
In such a case, the conservation of chiral currents $(J,G,\BG,T)$ in 
Eqs.(\ref{currentj}){--}(\ref{currentt}) 
is recovered and these currents generate $\CN = 2$ superconformal algebra.
In the IR limit $e^2 \to \infty$, the gauge fields are decoupled. 
Then we can evaluate the operator product expansion of 
these currents by free propagators. 
These represent $\CN =2$ superconformal algebra
with central charge $c = 3(m-n-1)$. (See Appendix B.)

We shall also show derivation of the Konishi anomaly from 
another viewpoint \cite{Hori:2001ax}. 
We want to evaluate one-loop effects in the lowest component 
$J_- \sim \sum_I \psi_-^I\Bpsi_-^I - \sum_A b_-^A\Bb_-^A + (2i/e^2)\sigma \partial_-\Bsigma$ of the current (\ref{Eq:current}). 
To begin with, we compute the following one-loop amplitudes with an
arbitrary operator $\CO$, 
\begin{eqnarray}
\biggl\langle \lnp\sum_I\psi_-^I(x_1)\Bpsi_-^I(x_2)\rnp \CO \biggr\rangle 
&=& - {i \over \pi} \sum_I Q_I \int d^2z {\langle v_+(z)\CO \rangle \over (x_1^- - z^-)(x_2^- - z^-)} , \\
\biggl\langle \lnp \sum_A b_-^A(x_1)\Bb_-^A(x_2)\rnp \CO \biggr\rangle
&=& -{i \over \pi} \sum_A q_A \int d^2z {\langle v_+(z)\CO \rangle \over (x_1^- - z^-)(x_2^- - z^-)}.  
\end{eqnarray}
These equations lead us to\footnote{Here we used 
\begin{eqnarray*}
\biggl\langle \lnp \biggl({\partial \over \partial x_1^+}\psi^I_-(x_1)\Bpsi^I_-(x_2) 
	+ \psi^I_-(x_1){\partial \over \partial x_2^+}\Bpsi^I_-(x_2) \biggr)\rnp \CO \biggr\rangle 
&=& -{i \over \pi} Q_I \int d^2z \biggl\{ {i\pi \delta^2(x_1 - z) \over x_2^- - z^-} + {i\pi \delta^2(x_2 - z) \over x_1^- - z^-} \biggr\}\langle v_+(z) \CO \rangle \\
= - Q_I \biggl\langle {v_+(x_1) - v_+(x_2) \over x_1^- - x_2^-} \CO \biggr\rangle 
&\sim& - Q_I \biggl\langle \biggl(\partial_-v_+(x_2) + {x_1^+ - x_2^+ \over x_1^- - x_2^-}\partial_+v_+(x_2) \biggr) \CO \biggr\rangle .
\end{eqnarray*}
The same method is applicable to the computation of 
$\langle \lnp \{\partial_{x_1^+}b_-^A(x_1)\Bb_-^A(x_2) +
b_-^A(x_1)\partial_{x_2^+}\Bb_-^A(x_2) \}\rnp \CO \rangle$. 
(See also Appendix B.)
}
\begin{eqnarray}
&&\partial_+ \biggl\langle \lnp \biggl(\sum_I \psi_-^I\Bpsi_-^I(x) - \sum_A b^A_- \Bb^A_-(x)
\biggr)\rnp \CO \biggr\rangle \nonumber\\
&&\qquad \sim \biggl(\sum_I Q_I - \sum_A q_A \biggr)\biggl\langle \biggl(\partial_-v_+(x) 
+ \lim_{x_1\to x_2} {x_1^+ - x_2^+ \over x_1^- -
  x_2^-}\partial_+v_+(x) \biggr) \CO\biggr\rangle . \label{Eq:1-loopamb}
\end{eqnarray}
Since there is ambiguity remained in Eq.(\ref{Eq:1-loopamb}), 
we should define the gauge invariant currents, 
\begin{eqnarray}
\psi_-^I\Bpsi_-^I(x) &\equiv& \lim_{x_1 \to x_2}
\left[\psi_-^I(x_1)\exp\left({iQ_I \over 2}\int_{x_2}^{x_1} dx^\mu v_\mu\right) \Bpsi_-^I(x_2) - {-i \over x_1^- - x_2^-}\right] \nonumber \\
&\sim& \lnp \psi^I_-\Bpsi^I_-(x) \rnp + {Q_I \over 2}
\lim_{x_1 \to x_2}{x_1^+ - x_2^+ \over x_1^- - x_2^-}v_+(x) 
+ {Q_I \over 2} v_-(x) , \label{Eq:ginvphiphiprod}\\
b_-^A\Bb_-^A(x) &\equiv& \lim_{x_1 \to x_2}
\left[b_-^A(x_1)\exp\left({iq_A \over 2}\int_{x_2}^{x_1} dx^\mu v_\mu\right) \Bb_-^A(x_2) - {-i \over x_1^- - x_2^-}\right] \nonumber \\
&\sim& \lnp b^A_-\Bb^A_-(x) \rnp + {q_A \over 2}
\lim_{x_1 \to x_2}{x_1^+ - x_2^+ \over x_1^- - x_2^-}v_+(x) 
+ {q_A \over 2} v_-(x) . \label{Eq:ginvbbprod}
\end{eqnarray}
The limit ``$\displaystyle \lim_{x_1\to x_2}$'' means that 
$x_1\to x$ and $x_2\to x$. 
Under these definitions, the chiral anomaly is correctly determined from Eq.(\ref{Eq:1-loopamb}) as
\begin{equation}
\biggl\langle \partial_+ \biggl(\sum_I \psi_-^I\Bpsi_-^I(x) 
	- \sum_A b_-^A \Bb_-^A(x) \biggr) \CO \biggr\rangle 
\sim \biggl(\sum_I Q_I - \sum_A q_A \biggr) \langle v_{+-}(x) \CO \rangle .
\end{equation}

Next we shall show (\ref{Eq:ginvphiphiprod}) and (\ref{Eq:ginvbbprod}) reproduce 
the Konishi anomaly. 
Since $\BD_+ J$ can be evaluated by $[\BQ_+, J]$, we obtain
\begin{equation}
\BD_+ J \sim \biggl(\sum_I Q_I - \sum_A q_A\biggr) \lambda_- , \label{Eq:Kanomaly1st} 
\end{equation} 
where we used the supersymmetric transformations 
$[\BQ_+, v_+]=0$ and $[\BQ_+, v_-]=2i\lambda_-$. 
The right hand side of Eq.(\ref{Eq:Kanomaly1st}) reminds us 
of the first component of $\BD_-\Sigma$, which is $\sqrt{2}i\lambda_-$. 
In fact, the supersymmetric completion of Eq.(\ref{Eq:Kanomaly1st}) becomes 
\begin{equation}
\BD_+ \CJ  \sim \biggl(\sum_I Q_I - \sum_A q_A\biggr)\BD_-\Sigma .
\end{equation}
That reproduces the results in the previous discussion and we can read that 
the vanishing condition of the Konishi anomaly is $\sum_I Q_I - \sum_A q_A = 0$.
In this case we have conformal currents in the IR region, 
which are identified with ${\cal J}$. 
We shall study such case in the next section.

%%%%%%%%%%
\section{$\CN=2$ Superconformal Algebra}

\newcommand{\gp}{G}
\newcommand{\gm}{\overline{G}}

We consider $\CN=2$ currents of our model in the Euclidean case. 
In the IR limit $e^2 \to \infty$, the gauge fields are decoupled and 
we use free field representations.
After appropriate redefinitions and Wick-rotations, 
we can write down a set of $\CN=2$ superconformal currents
$(J,\gp,\gm,T)$ in the Euclidean case
\begin{eqnarray*}
J &=& -\sum_I\Bpsi^I\psi^I-\sum_A\Bb^Ab^A+\Bsigma\partial\sigma\,,\\
\gp &=&-i\sqrt{2}\sum_I\psi^I\partial\Bphi^I+i\sqrt{2}\sum_Ab^A\partial\Bxi^A
-\sqrt{2}\Blambda \partial\sigma , \\
\gm &=&-i\sqrt{2}\sum_I\Bpsi^I\partial\phi^I+i\sqrt{2}\sum_A\Bb^A\partial\xi^A-\sqrt{2} \Bsigma\partial \lambda , \\
T &=& -\sum_I\partial\Bphi^I\partial\phi^I-\sum_A\partial\Bxi^A\partial\xi^A
-\Blambda\partial\lambda-\frac{1}{2}(\partial\Bsigma\partial\sigma-\Bsigma\partial^2\sigma) \\
&&-\frac{1}{2}\sum_I(\Bpsi^I\partial\psi^I-\partial\Bpsi^I\cdot\psi^I)
-\frac{1}{2}\sum_A(\Bb^A\partial b^A-\partial\Bb^A\cdot b^A) . 
\end{eqnarray*}
Here each field has operator product expansion
\begin{eqnarray*}
&&\Bphi^I(z)\phi^J(w)\sim -\delta^I{}^J\log (z-w) ,\quad 
\Bpsi^I(z)\psi^J(w)
\sim \frac{\delta^I{}^J}{z-w} , \\
&&\partial \Bxi^A(z)\xi^B(w)\sim \frac{\delta^A{}^B}{z-w} ,\quad 
\Bb^A(z)b^B(w)\sim \frac{-\delta^A{}^B}{z-w} , \\
&&\Blambda(z)\lambda (w)\sim \frac{1}{z-w} ,\quad 
\Bsigma(z)\partial\sigma (w)\sim \frac{1}{z-w} .
\end{eqnarray*}
Then we obtain $\CN=2$ superconformal algebra with 
central charge $c=3(m-n-1)$
\begin{eqnarray*}
&&T(z)T(w)\sim \frac{c/2}{(z-w)^4}+\frac{2}{(z-w)^2}T(w)+\frac{1}{z-w}\partial T(w) , \\
&&T(z)J(w)\sim \frac{1}{(z-w)^2}J(w)+\frac{1}{z-w}\partial J(w) , \\
&&T(z)\gp (w)\sim \frac{3/2}{(z-w)^2}\gp (w)+\frac{1}{z-w}\partial \gp (w) , \\
&&T(z)\gm (w)\sim \frac{3/2}{(z-w)^2}\gm (w)+\frac{1}{z-w}\partial \gm (w) , \\
&&J(z)\gp (w)\sim \frac{+1}{z-w}\gp (w) , \quad 
J(z)\gm (w)\sim \frac{-1}{z-w}\gm (w) , \\
&&\gp (z)\gm (w)\sim \frac{2c/3}{(z-w)^3}
+\frac{2}{(z-w)^2}J(w)+\frac{1}{z-w}(2T+ \partial J)(w) , \\
&&J(z)J(w)\sim \frac{c/3}{(z-w)^2} , \quad 
\gp (z)\gp (w)\sim 0 , \quad \gm (z)\gm (w)\sim 0 . 
\end{eqnarray*}
Conformal weights and $U(1)$ charges of fields are measured in terms of $T$ and $J$. 
Field $\Bsigma$ is the lowest component of the 
field strength of the vector superfield $\BSigma$. 
But this $\Bsigma$ has conformal weight $1/2$ 
with respect to $T$. 
Similarly the gaugino $\Blambda$ in the 
multiplet $\BSigma$ has weight $1$.
We summarize these data in Table~\ref{table1}.
\begin{table}
\[
\begin{array}{|c|c|c|}\hline
\mbox{fields} & \mbox{weights} & \mbox{$U(1)$ charges}\\\hline
(\psi^I,\Bpsi^I) & (1/2,1/2) & (+1,-1)\\
(b^A,\Bb^A) & (1/2,1/2) & (+1,-1)\\
(\partial\sigma ,\Bsigma) & (1/2,1/2) & (+1,-1)\\
(\lambda,\Blambda) & (0,1) & (0,0)\\
(\xi^A,\partial\Bxi^A) & (0,1) & (0,0)\\\hline
\end{array}
\]
\caption{Conformal weights and $U(1)$ charges of fields}\label{table1}
\end{table}
By taking account of the conformal weights, we can
redefine fields $\sigma ,\Bsigma, \lambda ,\Blambda$ 
by $b^{\sigma},\Bb^{\sigma},\xi^{\sigma},\Bxi^{\sigma}$
\begin{eqnarray*}
&&b^{\sigma}=i\partial\sigma , \quad \Bb^{\sigma}=i\Bsigma , \quad
\xi^{\sigma}=\lambda , \quad \partial\Bxi^{\sigma}=\Blambda , \\
&&\Phi^P=(\phi^I, \xi^A, \xi^{\sigma} ; \Bphi^I, \Bxi^A, \Bxi^{\sigma})
	=(\Phi^p; \Phi^{\overline{p}}) , \\
&&\Psi^P=(\psi^I, b^A, b^{\sigma};\Bpsi^I, \Bb^A, \Bb^{\sigma})
	=(\Psi^p; \Psi^{\overline{p}}) .
\end{eqnarray*}
Also we introduce symbol $|P|$ $(P=I,A,\sigma)$ as
$|I|=0$ and $|A|=|\sigma |=1$.
Then $\Phi^P$ and $\Psi^P$ have commutation or anti-commutation
relations
as follows: 
\begin{eqnarray*}
\Phi^P\cdot \Phi^Q &=& (-1)^{|P||Q|}\Phi^Q\cdot \Phi^P , \\
\Psi^P\cdot \Psi^Q &=& -(-1)^{|P||Q|}\Psi^Q\cdot \Psi^P , \\
\Phi^P\cdot \Psi^Q &=& (-1)^{|P|(|Q|+1)}\Psi^Q\cdot \Phi^P .
\end{eqnarray*}
Under this setup,  $\CN=2$ superconformal currents are expressed 
by $\Phi^P$ and $\Psi^P$ 
\begin{eqnarray*}
J &=& -\frac{1}{2}N_{PQ}\Psi^P{\Psi}^Q ,\\
\gp &=& -i\sqrt{2}\cdot\frac{1}{2}(M-N)_{PQ}\Psi^P\partial {\Phi}^Q , \\
\gm &=&-i\sqrt{2}{\Psi}^P\partial {\Phi}^Q\cdot \frac{1}{2}(M-N)_{QP} , \\
T &=& -\frac{1}{2}M_{PQ}\partial\Phi^P\partial {\Phi}^Q
-\frac{1}{2}M_{PQ}\Psi^P\partial {\Psi}^Q , 
\end{eqnarray*}
where $M_{PQ}$ and $N_{PQ}$ have non-zero components
\begin{eqnarray*}
&&M_{I\BJ}=M_{\BJ I}=\delta_{IJ} , \quad 
M_{A\overline{B}}=-M_{\overline{B} A}=-\delta_{AB} , \quad 
M_{\sigma\Bsigma}=-M_{\Bsigma\sigma}=-1 , \\
&&N_{I\BJ}=-N_{\BJ I}=-\delta_{IJ}, \quad
N_{A\overline{B}}=N_{\overline{B} A}=\delta_{AB}, \quad
N_{\sigma\Bsigma}=N_{\Bsigma\sigma}=1. 
\end{eqnarray*}
Also we can introduce their inverse matrices $M^{PQ}$ and $N^{PQ}$ with 
$M_{PR}M^{RQ}=\delta_P{}^Q$ and $N_{PR}N^{RQ}=\delta_P{}^Q$. 
Propagators of fields $\Phi^P$ and $\Psi^P$ are expressed as 
\[
\Phi^P(z)\Phi^Q(w)\sim -M^{PQ}\log (z-w) , \quad 
\Psi^P(z)\Psi^Q(w)\sim \frac{M^{PQ}}{z-w} . 
\]

In order to analyze geometric picture of manifolds, 
we consider twisted conformal field theories by introducing 
modified energy momentum tensors $\tilde{T}^{(\pm)}=T\pm \frac{1}{2}\partial J$
\[
\tilde{T}^{(\pm)}=-\frac{1}{2}M_{PQ}\partial\Phi^P\partial {\Phi}^Q-
\frac{1}{2}(M\pm N)_{PQ}\Psi^P\partial {\Psi}^Q . 
\]
When one measures conformal weights $\tilde{h}$ of fields 
by these new energy momentum tensors, 
$\Psi^p$ $(p=I,A,\sigma)$ and 
${\Psi}^{\overline{p}}$ $(\overline{p}=\overline{I},\overline{A},\Bsigma)$ 
have $\tilde{h}=0,1$ for $\tilde{T}^{(+)}$, 
$\tilde{h}=1,0$ for $\tilde{T}^{(-)}$. 
Also super stress tensors $\gp$ ($\gm$) have respectively 
conformal weights $1$ under $\tilde{T}^{(+)}$ $(\tilde{T}^{(-)})$.
So we can define BRST operators for these twisted models
\begin{eqnarray*}
Q = -\frac{i}{\sqrt{2}}\oint \gp &\quad& \mbox{for $\tilde{T}^{(+)}$} , \\
\BQ = -\frac{i}{\sqrt{2}}\oint \gm &\quad& \mbox{for $\tilde{T}^{(-)}$} .
\end{eqnarray*}
These BRST charges act on fields 
with parameters $\alpha$, $\tilde{\alpha}$ in the following way and 
they can be interpreted as some kinds 
of ``differential operators'' on the supermanifold
\begin{eqnarray*}
&&[\alpha Q,\Phi^p]= \alpha \Psi^p , \quad
[\tilde{\alpha} \BQ,\Phi^{\overline{p}}]= \tilde{\alpha} \Psi^{\overline{q}}
S_{\overline{q}}{}^{\overline{p}} ,\\
&&[\alpha Q,\Psi^{\overline{p}}]=-\alpha \partial \Phi^{\overline{q}}
S_{\overline{q}}{}^{\overline{p}} , \quad
[\tilde{\alpha} \BQ,\Psi^p]=-\tilde{\alpha} \partial \Phi^p , \\
&&S=
\left(
\begin{array}{cc}
+I_m & 0 \\
0 & -I_{n+1}
\end{array}
\right) . 
\end{eqnarray*}
In this setting, $\Psi^p$ $(p=I,A,\sigma)$ and 
${\Psi}^{\overline{p}}$ $(\overline{p}=\overline{I},\overline{A},\Bsigma)$ are naively set to 
differential forms on the manifold 
\begin{eqnarray*}
(d\phi^I\quad d\xi^A\quad d\xi^{\sigma}) &\leftrightarrow& (\psi^I\quad b^A\quad b^{\sigma}) , \\
(d\Bphi^I\quad d\Bxi^A\quad d\Bxi^{\sigma})
&\leftrightarrow& (\Bpsi^I\quad \Bb^A\quad \Bb^{\sigma}) . 
\end{eqnarray*}
For the purpose of further investigation, let us bosonize fields 
in $\Psi^P$
\begin{eqnarray*}
&&\psi^I=e^{\varphi^I} ,\quad \Bpsi^I=e^{-\varphi^I} , \\
&&b^A=e^{\varphi^A}\eta^A ,\quad \Bb^A=e^{-\varphi^A}\partial\zeta^A , \\
&&b^{\sigma}=e^{\varphi^{\sigma}}\eta^{\sigma} , \quad 
	\Bb^{\sigma}=e^{-\varphi^{\sigma}}\partial\zeta^{\sigma} , \\
&&\varphi^I(z)\varphi^J(w)\sim \delta^{IJ}\log (z-w) , \\
&&\varphi^A(z)\varphi^B(w)\sim -\delta^{AB}\log (z-w) , \quad
\zeta^A(z)\eta^B(w)\sim \frac{\delta^A{}^B}{z-w} , \\
&&\varphi^{\sigma}(z)\varphi^{\sigma}(w)\sim -\log (z-w) ,\quad
\zeta^{\sigma}(z)\eta^{\sigma}(w)\sim \frac{1}{z-w} . 
\end{eqnarray*}
Then $(\psi^I,\Bpsi^I)$ can be identified with $(b,c)$ systems with spins $(1,0)$. 
On the other hand, $(b^A,\Bb^A)$ and $(b^{\sigma},\Bb^{\sigma})$ 
correspond to ``bosonic ghost'' systems $(\beta,\gamma)$ with spins $(1,0)$. 
Our $\CN=2$ algebra contains $U(1)$ current $J$ that is the 
number current of these ghost and bosonic ghost fields in the context of 
twisted models.  Differential forms on the manifold can be 
represented as ghost fields in the twisted model.

Let us return to the untwisted model.
This $U(1)$ current is rewritten under the bosonization
\begin{eqnarray*}
&&J=\partial (\varphi^I-\varphi^A-\varphi^{\sigma})
=i\sqrt{\hat{c}}\,\partial H , \quad 
\hat{c}=m-n-1 , \\
&&H(z)H(w)\sim -\log (z-w) . 
\end{eqnarray*}
This formula means that 
the $U(1)$ current $J$ is the sum of 
fermion number current $J_F=-\Bpsi^I\psi^I=\partial \varphi^I$ 
and ghost number current $J_P=-\partial (\varphi^A+\varphi^{\sigma})$.
The set of currents $(T,\gp ,\gm ,J)$ generate 
$\CN=2$ superconformal algebra with central charge $c=3(m-n-1)$. 
This algebra has algebra automorphism and NS sector and R sector are 
related by a spectral flow operator.
Under this flow, the $(h,q)=(0,0)$ state is transformed into 
states associated with primary fields $\Omega$ and $\BOmega$ 
\begin{eqnarray*}
&&\Omega =e^{i\sqrt{\hat{c}}H}
=e^{\varphi^I-\varphi^A-\varphi^{\sigma}}=e^{\varphi^I}\delta (b^A)\delta (b^{\sigma}) , \\
&&\BOmega=e^{-i\sqrt{\hat{c}}H}
 =e^{-\varphi^I+\varphi^A+\varphi^{\sigma}}=e^{\overline{\varphi}^I}
\delta (\Bb^A)\delta (\Bb^{\sigma}) . 
\end{eqnarray*}
These operators satisfy chiral primary condition $h=q/2$ with  
$q=\hat{c}$ for $\Omega$, $q=-\hat{c}$ for $\BOmega$.
By looking at this formula, we could identify 
$d\xi^A$, $d\xi^{\sigma}(=d\lambda)$ 
with $\delta (b^A)$, $\delta (b^{\sigma})$ respectively.
It means that the degree of differential forms on the supermanifold 
is measured by $U(1)$ charges associated with the current $J$, and 
differential forms are described 
by $\psi^I$'s, $\delta (b^A)$'s, $\delta (b^{\sigma})$. 
In the geometric picture of the sigma model, $\Omega$ and $\BOmega$ 
are respectively associated with the holomorphic $\hat{c}$ form 
and the antiholomorphic $\hat{c}$ form. 

Next we shall consider superpartners $U$, $\BU$ of $\Omega$, $\BOmega$.
They are defined by considering operator products
\begin{eqnarray*}
&&\gm (z)\Omega (w)\sim \frac{i\sqrt{2}}{z-w}U(w) , \quad 
\gp (z)\BOmega(w)\sim \frac{i\sqrt{2}}{z-w}\BU(w) , \\ 
&&U=-\sum_I\exp\biggl(\sum_{J\neq I}\varphi^J-\sum_A\varphi^A-\varphi^{\sigma}\biggr)\partial\phi^I \\
&&\qquad +\sum_A\exp\biggl(-2\varphi^A+\sum_{I}\varphi^I-\sum_{B\neq A}
\varphi^B-\varphi^{\sigma}\biggr)\partial\zeta^A\partial\xi^A \\
&&\qquad +\exp\biggl(-2\varphi^{\sigma}+\sum_{I}\varphi^I
-\sum_A\varphi^A\biggr)\partial\zeta^{\sigma}\partial\xi^{\sigma} , \\
&&\BU=-\sum_I\exp\biggl(-\sum_{J\neq I}\varphi^J+\sum_A\varphi^A+\varphi^{\sigma}\biggr)\partial\Bphi^I \\ 
&&\qquad +\sum_A\exp\biggl(2\varphi^A-\sum_{I}\varphi^I+\sum_{B\neq A}\varphi^B+\varphi^{\sigma}\biggr)
\eta^A\partial\Bxi^A \\
&&\qquad +\exp\biggl(2\varphi^{\sigma}-\sum_{I}\varphi^I+\sum_A\varphi^A\biggr)
\eta^{\sigma}\partial\Bxi^{\sigma} ,
\end{eqnarray*}
where we omit cocycle factors for each term.
These superpartners are primary fields with $h=(\hat{c}+1)/2$ 
and $U$, $\BU$ have $U(1)$ charges $q=\hat{c}-1,-(\hat{c}-1)$ respectively (see Table~\ref{table2}). 
\begin{table}
\[
\begin{array}{|c|c|c|c|c|}\hline
\mbox{fields} & \Omega & \BOmega & U & \BU \\ \hline
\mbox{charge $q$} & \hat{c} & -\hat{c} & \hat{c}-1 & -(\hat{c}-1)\\ \hline
\mbox{weight $h$} & \hat{c}/2 & \hat{c}/2 & (\hat{c}+1)/2 & (\hat{c}+1)/2 \\ \hline
\end{array}
\]
\caption{Conformal weights and $U(1)$ charges of fields $\Omega$, $\BOmega$, 
$U$ and $\BU$.}\label{table2}
\end{table}
Together with these $U$, $\BU$, $\Omega$ and $\BOmega$, 
the $\CN=2$ superconformal algebra is enlarged to extended algebra. 
For $\hat{c}=1$ case, 
$\Omega$ and $\BOmega$ have conformal weight $1/2$ and $q=\pm 1$. 
They turn out to be a pair of complex fermions. 
On the other hand, neutral fields $U$ and $\BU$ have 
$h=1$ and they are interpreted as derivatives of scalars, namely  
these fields correspond to a pair of complex bosons.
Resulting algebra is 
direct product of $\CN =2$ superconformal algebra with $\hat{c}=1$ and 
a pair of complex bosons and fermions. 
Let us see cases of $\hat{c}=2,3$ concretely.

\subsection{$\hat{c}=3$ case}

\begin{figure}
\begin{center}
\input{n2.tex}
\end{center}
\caption{Currents of $c=9$ extended algebra. 
In this figure, $h$ is the conformal weight and $q$ is the $U(1)$ charge measured by $J$. }\label{fig1}
\end{figure}
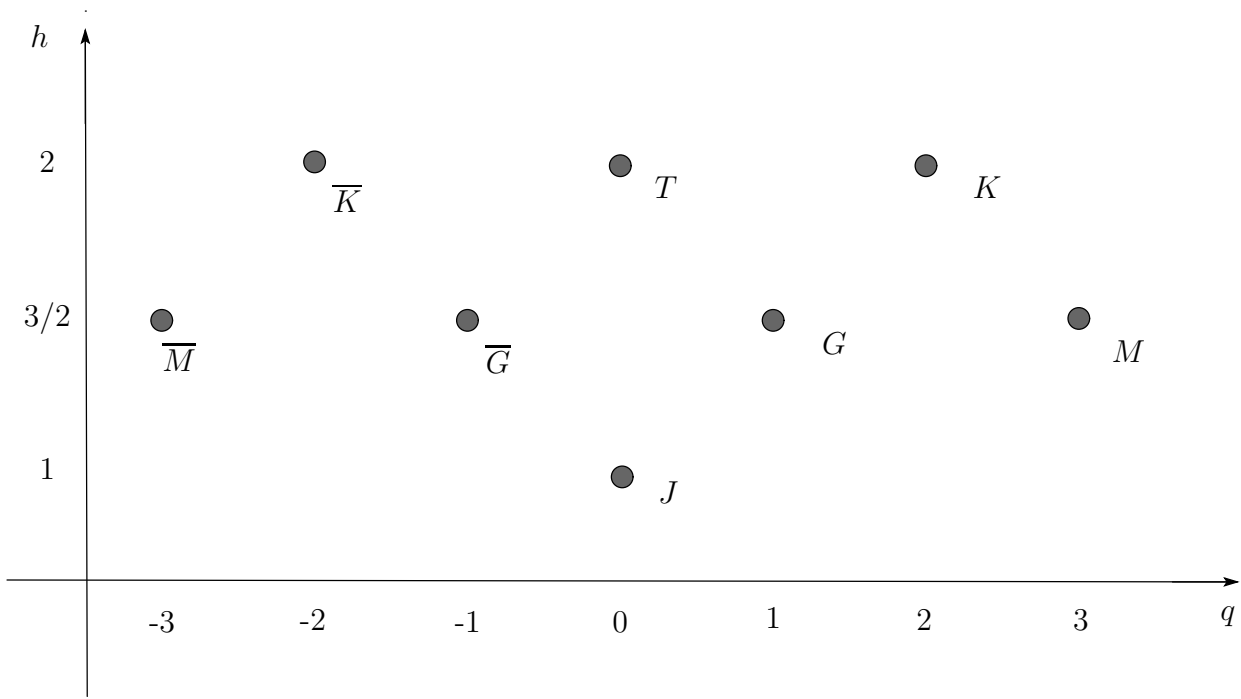

This case corresponds to a Calabi-Yau threefold 
and resulting world-sheet theory is 
described by $c=9$ algebra. 
When we put 
$M=\sqrt{2}\Omega$, $\overline{M}=\sqrt{2}\,\BOmega$, 
$K=iU$, $\overline{K}=i\BU$, the extended algebra is 
expanded by eight currents $(T,\gp ,\gm ,J, M,\overline{M}, K, \overline{K})$. 
They are shown in Fig.~\ref{fig1}. 
$K$ and $\overline{K}$ are respectively superpartners 
of $\Omega$ and $\BOmega$. 
They contain subalgebra with $\CN =2$ superconformal symmetry with
$\hat{c}=1/3$ $(c=1)$. It is generated by currents 
$(\hat{T},\hat{\gp},\overline{\hat G},\hat{J})$
\begin{eqnarray*}
&&J=\partial \left(\sum_I\varphi^I-\sum_A\varphi^A-\varphi^{\sigma}\right) , \\
&&\hat{T}=\frac{1}{6}J^2 , \quad \hat{J}=\frac{1}{3}J , \quad 
\hat{\gp}=\sqrt{\frac{2}{3}}\Omega , \quad \overline{\hat G}=\sqrt{\frac{2}{3}}\BOmega . 
\end{eqnarray*}
This conformal subalgebra belongs to the $\CN =2$ minimal model and its 
spectrum is classified by 
$(\hat{h},\hat{q})=(0,0), (1/6,\pm 1/3)$. In other words, these states are 
labeled by the original $U(1)$ charge $q=0,\pm 1$. 
The character of this subalgebra is expressed 
by the Dedekind's eta-function $\eta (\tau)$ 
and classical SU(2) theta function $\Theta_{m,k}(\tau ,\theta)$ as
$\chi_{q}=\eta^{-1} (\tau) \Theta_{2q,3}(\tau/2,\theta)$.

\subsection{$\hat{c}=2$ case}

\begin{figure}
\begin{center}
\input{n4.tex}
\end{center}
\caption{Currents of $\hat{c}=2$ theory. They generate $\CN=4$ superconformal algebra 
with affine $\widehat{su}(2)_1$ algebra. In this figure, $h$ is the conformal weight and $q$ is the 
$U(1)$ charge measured by $J$. 
The $su(2)$ current $J^3$ is related with $J$ through a relation $J^3=J/2$.}\label{fig2}
\end{figure}
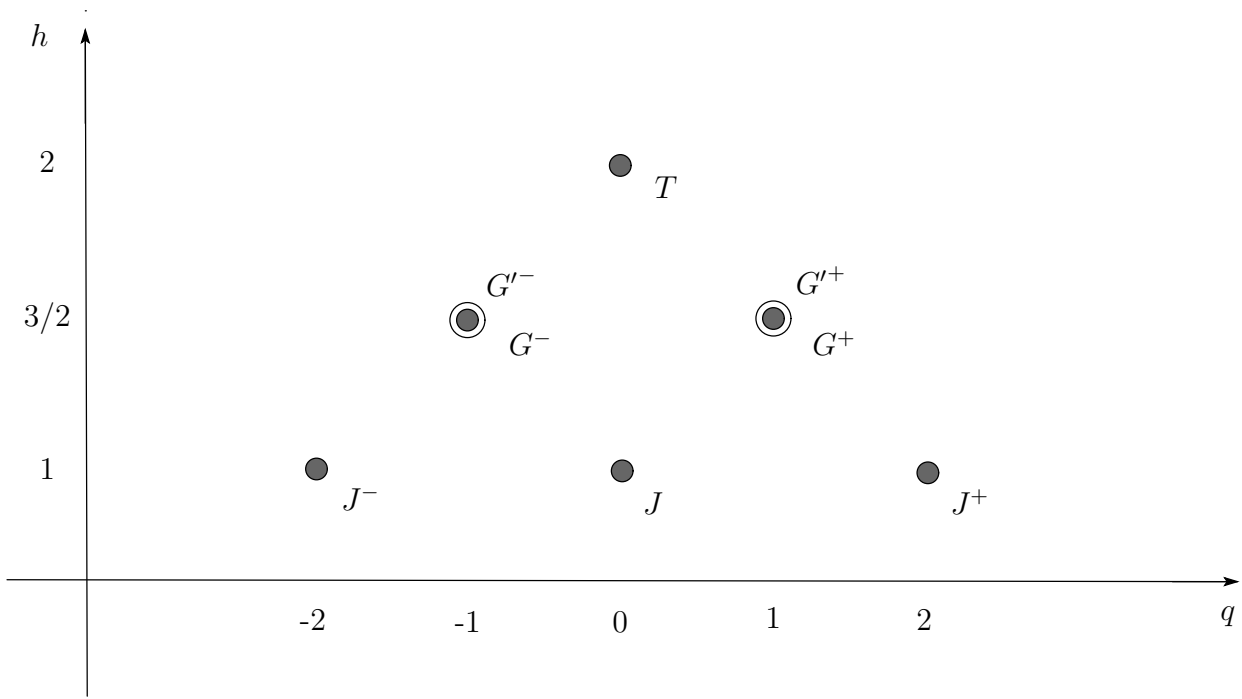

This case is an analogue of four-dimensional hyperk\"ahler case and  
resulting world-sheet theory is described by 
$\CN =4$ superconformal algebra. 
When we put 
\begin{eqnarray*}
&&J^+=i\Omega , \quad J^-=-i\BOmega , \quad J^3=\frac{1}{2}J , \\
&&G^+=G , \quad G^-=\BG , \quad G'^+=\sqrt{2}U , \quad 
G'^-=\sqrt{2}\BU , 
\end{eqnarray*}
eight currents $(T, G^{\pm}, {G'}^{\pm}, J^{\pm}, J)$ 
generate $\CN=4$ superconformal 
algebra with $c=6$. It includes affine $\widehat{su}(2)_1$ algebra which 
is constructed by $J^{\pm}$ and $J^3= J/2$. 
This symmetry seems to reflect an analogue 
of the hyperk\"ahler structure of the 
manifold.  Choice of the complex structure corresponds to pick up the 
K\"ahker class associated to the $U(1)$ current $J$. 
The other almost complex structures are related to 
$(2,0)$-form $\Omega$ and $(0,2)$-form $\BOmega$. 
Also  $(G^+,{G'}^-)$ and $({G'}^+,G^-)$ turn out to be 
doublets under this $su(2)$. 
Two components in each doublet are changed one another 
under the action of $J^{\pm}$. 
We summarize conformal weight $h$ and $U(1)$ charge $q$ 
in Fig.~\ref{fig2}.
 Here we show charge $q$ measured by $J$.  
The $su(2)$ current $J^3$ is related to this $J$ through $J^3=J/2$.

\section{Conclusions}

We have studied the gauged linear sigma model on the supermanifold with 
Grassmann odd chiral superfields, 
which provide Grassmann odd coordinates in the target supermanifold. 
In this paper we have considered the $U(1)$ gauge theory and 
target space of this model is reduced to the weighted projective space 
$\mathbb{WCP}^{m-1|n}$. 
This model has supersymmetry on the world-sheet and 
there are classically conserved $U(1)$ R-symmetries.
If these R-symmetries are not anomalous, 
the resulting theory  has superconformal symmetry in the IR region. 
But conservation of supercurrents could be broken generally due to 
anomalies for these $U(1)$ symmetries. 

In order to study quantum properties of the theory, we have 
calculated the Konishi anomaly associated with the R-symmetries. 
We have taken two approaches in this calculation; 
covariant calculation by the path integral methods and 
evaluation of one-loop effects in the light-cone gauge. 
By these analyses, the condition of anomaly cancellation 
$\sum_IQ_I-\sum_Aq_A=0$ is obtained. 
The result is consistent with the Ricci-flat condition of the 
supermanifold. If this condition is satisfied, then 
the theory flows into a superconformal theory in the IR limit. 
We also constructed superconformal currents explicitly in this region, 
and found that they generate $\CN=2$ 
superconformal algebra with central charge $c=3(m-n-1)$. 

In this model, there are spin $1/2$ fields, 
$(\psi^I,\Bpsi^I)$ and $(b^A,\Bb^A)$ that 
contribute $+m$ and $-n$ to the central charge of this algebra. 
The lowest components $\sigma$ and $\Bsigma$ of field strengths 
$\Sigma$ and $\BSigma$
of vector superfield play essential roles in this 
model. The set $(\partial\sigma ,\Bsigma)$ has conformal weights 
$(1/2,1/2)$ respectively and 
behaves as an extra set of $(b^A,\Bb^A)$'s. Especially it induces extra contribution ``$-1$'' in 
the formula of $c$.  

These fields are collected into the $U(1)$ current of $\CN =2$ superconformal algebra. 
By bosonizing the bosonic fields, we find that 
the $U(1)$ current $J$ is the sum of 
the fermion number current $J_F=\psi^I\Bpsi^I$ and 
ghost number current $J_P=-\partial (\varphi^A+\varphi^{\sigma})$. 
This $J$ measures ``charges'' of fields, that are also identified with 
degrees of differential forms on the manifolds. 
Usually fermions $\psi^I$ and $\Bpsi^I$ are interpreted as 
differential forms on the manifold and 
charges associated with $J_F$ are identified with degrees of these forms.  
In our supermanifolds, bosonic fields $b^A$,$\Bb^A$,$b^\sigma$, $\Bb^{\sigma}$ 
could be identified with differential forms of Grassmann odd coordinates. From 
the correspondence between $U(1)$ charges and 
degrees of forms, we suppose that ghost number is identified with 
degree of differential forms of Grassmann odd coordinates.

In order to put forwards this interpretation, we 
discussed the holomorphic form $\Omega$ and 
its conjugate. Their expression has 
terms $\delta (b^A)\delta (b^{\sigma})$ and their conjugates that 
confirm the above supposition.

Finally we investigate structure of extended algebra generated by 
these $\Omega, \BOmega$ and their superpartners. 
The $\hat{c}=3$ case is an analogue of the Calabi-Yau threefold and 
resulting theory is described by $c=9$ algebra on the world-sheet. 
For $\hat{c}=2$ case, the algebra is enlarged into $\CN=4$ superconformal 
symmetry with affine $\widehat{su}(2)_1$. 
This model is an analogue of four-dimensional hyperk\"ahler manifolds. 
It implies that the affine $su(2)$ reflects some kind of 
hyperk\"ahler structure of the manifold. 
We are looking forward to further investigation towards this direction.

%%%%%%%%%%
\bigbreak\bigskip\bigskip
\centerline{{\bf Acknowledgments}}\nobreak

The research of S.S. was supported in part by the Grant-in-Aid 
for Scientific Research (\#16740159) from the Ministry 
of Education, Culture, Sports, Science and Technology (MEXT) of Japan 
and also by the Grant-in-Aid for the 21st Century COE 
``Center for Diversity and Universality in Physics'' from MEXT.
%%%%%%%%%%
\bigbreak\bigskip\bigskip

\appendix
\section{Conventions}

\begin{eqnarray}
&&x^\pm \equiv x^0 \pm x^1, \quad
\partial_\pm \equiv {1 \over 2}(\partial_0 \pm \partial_1) ,\\
&&\partial_+\partial_- \ln |x^+x^-| = i \pi \delta(x^0)\delta(x^1) 
	\equiv i \pi \delta^2(x), \\
&&\partial_- {1 \over x^+} = i \pi \delta^2(x),\quad 
\partial_+ {1 \over x^-} = i \pi \delta^2(x) , \\
&&\int d^2 \theta \equiv {1 \over 2} \int d\theta^- d\theta^+, \quad 
\int d^2 \Btheta \equiv {1 \over 2} \int d\Btheta^+ d\Btheta^-, \\
&&\int d^4 \theta \equiv \int d^2\theta d^2 \Btheta 
= {1 \over 4} {\partial \over \partial \Btheta^+}{\partial \over \partial \Btheta^-}{\partial \over \partial \theta^-}{\partial \over \partial \theta^+} .
\end{eqnarray}

\section{Superconformal Algebra (Lorentzian case)}

\subsection{Two point functions}
\begin{eqnarray}
&&\Bphi^I(x) \phi^J(y) \sim -{\delta^{IJ} \over 2} \ln |x-y|^2, \quad 
\Bpsi_\pm^I(x) \psi_\pm^J(y) \sim -{i \delta^{IJ} \over x^\pm - y^\pm}, \label{Eq:phphope}\\
&&\Bxi^A(x) \xi^B(y) \sim {\delta^{AB} \over 2} \ln |x-y|^2, \quad
\Bb_\pm^A(x) b_\pm^B(y) \sim {i \delta^{AB} \over x^\pm - y^\pm} , \\ 
&&\Blambda_\pm(x) \lambda_\pm(y) \sim -i e^2 {1 \over x^\pm - y^\pm} , \quad 
\Bsigma(x) \sigma(y) \sim -{e^2 \over 2} \ln |x-y|^2, \label{Eq:sisiope}\\
&&v_\pm(x) v_\pm(y) \sim \biggl(-{e^2 \over 8} + {\alpha \over 2}\biggr)
	{x^\mp - y^\mp \over x^\pm - y^\pm} ,\quad 
v_\pm(x) v_\mp(y) \sim \biggl({e^2 \over 8} + {\alpha \over 2}\biggr)
	\ln |x - y|^2 . 
\end{eqnarray}
We added $-{1 \over 2\pi} \int d^2y{1 \over 8\alpha}(\partial^\mu v_\mu)^2$ 
to the action for gauge fixing.

\subsection{$\CN=2$ superconformal algebra}

Superconformal currents are expressed by Eqs.(\ref{Eq:J1st}){--}(\ref{Eq:J4th}).
When one considers the IR limit $e^2 \to \infty$, 
the gauge fields are decoupled. 
Then we can evaluate the operator product expansion of 
these currents by the free propagators (\ref{Eq:phphope}){--}(\ref{Eq:sisiope}), 
\begin{eqnarray}
J(x)J(y) &\sim& -{m-n-1 \over (x^--y^-)^2} , \label{Eq:JJope} \\
J(x)G(y) &\sim& {i \over x^--y^-}G(y) , \\
J(x)\BG (y) &\sim& {-i \over x^--y^-}\BG (y) , \\
T(x)J(y) &\sim& {J(y) \over (x^--y^-)^2} + {\partial_-J(y) \over x^--y^-}, \\
T(x)G(y) &\sim& {3G(y) \over 2(x^--y^-)^2} + {\partial_-G(y) \over x^--y^-} , \\
T(x)\BG (y) &\sim& {3\BG (y) \over 2(x^--y^-)^2} + {\partial_- \BG (y) \over x^--y^-} , \\
T(x)T(y) &\sim& {3(m-n-1) \over 2(x^--y^-)^4} 
	+ {2T(y) \over (x^--y^-)^2} + {\partial_-T(y) \over x^--y^-} , \\
G(x)\BG (y) &\sim& {2i(m-n-1) \over (x^--y^-)^3} + {2J(y) \over (x^--y^-)^2}
	+{2iT(y) + \partial_-J_-(y) \over x^--y^-} . 
\end{eqnarray}
These relations represent $\CN=2$ superconformal algebra
with central charge $c = 3(m-n-1)$.

%%%%%%%%%%%%%%%%%%%%%%%%
%%%%%  REFERENCES  %%%%%
%%%%%%%%%%%%%%%%%%%%%%%%

%%%%%%%%%%%%%%%%%%%%%%

\end{document}

%% file: n2.tex
%WinTpicVersion3.08
\unitlength 0.1in
\begin{picture}( 64.4000, 35.9500)(  3.9000,-37.9000)
% CIRCLE 2 0 0 0
% 4 3600 1010 3560 1050 3560 1050 3590 1020
% 
\special{pn 8}%
\special{sh 0.600}%
\special{ar 3600 1010 58 58  0.0000000 6.2831853}%
% CIRCLE 2 0 0 0
% 4 4400 1820 4360 1860 4360 1860 4390 1830
% 
\special{pn 8}%
\special{sh 0.600}%
\special{ar 4400 1820 58 58  0.0000000 6.2831853}%
% CIRCLE 2 0 0 0
% 4 2800 1820 2760 1860 2760 1860 2790 1830
% 
\special{pn 8}%
\special{sh 0.600}%
\special{ar 2800 1820 58 58  0.0000000 6.2831853}%
% CIRCLE 2 0 0 0
% 4 3610 2640 3570 2680 3570 2680 3600 2650
% 
\special{pn 8}%
\special{sh 0.600}%
\special{ar 3610 2640 58 58  0.0000000 6.2831853}%
% CIRCLE 2 0 0 0
% 4 5200 1010 5160 1050 5160 1050 5190 1020
% 
\special{pn 8}%
\special{sh 0.600}%
\special{ar 5200 1010 58 58  0.0000000 6.2831853}%
% CIRCLE 2 0 0 0
% 4 6000 1810 5960 1850 5960 1850 5990 1820
% 
\special{pn 8}%
\special{sh 0.600}%
\special{ar 6000 1810 58 58  0.0000000 6.2831853}%
% CIRCLE 2 0 0 0
% 4 2000 990 1960 1030 1960 1030 1990 1000
% 
\special{pn 8}%
\special{sh 0.600}%
\special{ar 2000 990 58 58  0.0000000 6.2831853}%
% CIRCLE 2 0 0 0
% 4 1200 1820 1160 1860 1160 1860 1190 1830
% 
\special{pn 8}%
\special{sh 0.600}%
\special{ar 1200 1820 58 58  0.0000000 6.2831853}%
% LINE 2 0 3 0
% 4 390 3180 6780 3190 6780 3190 6780 3190
% 
\special{pn 8}%
\special{pa 390 3180}%
\special{pa 6780 3190}%
\special{fp}%
\special{pa 6780 3190}%
\special{pa 6780 3190}%
\special{fp}%
% LINE 2 0 3 0
% 2 810 3790 800 320
% 
\special{pn 8}%
\special{pa 810 3790}%
\special{pa 800 320}%
\special{fp}%
% VECTOR 2 0 3 0
% 2 6490 3190 6830 3190
% 
\special{pn 8}%
\special{pa 6490 3190}%
\special{pa 6830 3190}%
\special{fp}%
\special{sh 1}%
\special{pa 6830 3190}%
\special{pa 6764 3170}%
\special{pa 6778 3190}%
\special{pa 6764 3210}%
\special{pa 6830 3190}%
\special{fp}%
% VECTOR 2 0 3 0
% 2 800 200 800 200
% 
\special{pn 8}%
\special{pa 800 200}%
\special{pa 800 200}%
\special{fp}%
% VECTOR 2 0 3 0
% 2 800 790 800 300
% 
\special{pn 8}%
\special{pa 800 790}%
\special{pa 800 300}%
\special{fp}%
\special{sh 1}%
\special{pa 800 300}%
\special{pa 780 368}%
\special{pa 800 354}%
\special{pa 820 368}%
\special{pa 800 300}%
\special{fp}%
% STR 2 0 3 0
% 3 3600 3300 3600 3400 5 0
% 0
\put(36.0000,-34.0000){\makebox(0,0){0}}%
% STR 2 0 3 0
% 3 4400 3280 4400 3380 5 0
% 1
\put(44.0000,-33.8000){\makebox(0,0){1}}%
% STR 2 0 3 0
% 3 5190 3290 5190 3390 5 0
% 2
\put(51.9000,-33.9000){\makebox(0,0){2}}%
% STR 2 0 3 0
% 3 6010 3290 6010 3390 5 0
% 3
\put(60.1000,-33.9000){\makebox(0,0){3}}%
% STR 2 0 3 0
% 3 2800 3300 2800 3400 5 0
% -1
\put(28.0000,-34.0000){\makebox(0,0){-1}}%
% STR 2 0 3 0
% 3 1990 3290 1990 3390 5 0
% -2
\put(19.9000,-33.9000){\makebox(0,0){-2}}%
% STR 2 0 3 0
% 3 1200 3300 1200 3400 5 0
% -3
\put(12.0000,-34.0000){\makebox(0,0){-3}}%
% STR 2 0 3 0
% 3 600 2500 600 2600 5 0
% 1
\put(6.0000,-26.0000){\makebox(0,0){1}}%
% STR 2 0 3 0
% 3 600 1710 600 1810 5 0
% 3/2
\put(6.0000,-18.1000){\makebox(0,0){3/2}}%
% STR 2 0 3 0
% 3 600 890 600 990 5 0
% 2
\put(6.0000,-9.9000){\makebox(0,0){2}}%
% STR 2 0 3 0
% 3 6780 3270 6780 3370 5 0
% $q$
\put(67.8000,-33.7000){\makebox(0,0){$q$}}%
% STR 2 0 3 0
% 3 560 230 560 330 5 0
% $h$
\put(5.6000,-3.3000){\makebox(0,0){$h$}}%
% STR 2 0 3 0
% 3 3780 1080 3780 1180 2 0
% $T$
\put(37.8000,-11.8000){\makebox(0,0)[lb]{$T$}}%
% STR 2 0 3 0
% 3 3850 2620 3850 2720 5 0
% $J$
\put(38.5000,-27.2000){\makebox(0,0){$J$}}%
% STR 2 0 3 0
% 3 4720 1840 4720 1940 5 0
% $G$
\put(47.2000,-19.4000){\makebox(0,0){$G$}}%
% STR 2 0 3 0
% 3 2960 1920 2960 2020 5 0
% $\bar{G}$
\put(29.6000,-20.2000){\makebox(0,0){$\overline{G}$}}%
% STR 2 0 3 0
% 3 6260 1880 6260 1980 5 0
% $M$
\put(62.6000,-19.8000){\makebox(0,0){$M$}}%
% STR 2 0 3 0
% 3 5520 1020 5520 1120 5 0
% $K$
\put(55.2000,-11.2000){\makebox(0,0){$K$}}%
% STR 2 0 3 0
% 3 1290 1910 1290 2010 5 0
% $\bar{M}$
\put(12.9000,-20.1000){\makebox(0,0){$\overline{M}$}}%
% STR 2 0 3 0
% 3 2170 1080 2170 1180 5 0
% $\bar{K}$
\put(21.7000,-11.8000){\makebox(0,0){$\overline{K}$}}%
\end{picture}%

%% file: n4.tex
%WinTpicVersion3.08
\unitlength 0.1in
\begin{picture}( 64.4000, 35.9500)(  3.9000,-37.9000)
% CIRCLE 2 0 0 0
% 4 3600 1010 3560 1050 3560 1050 3590 1020
% 
\special{pn 8}%
\special{sh 0.600}%
\special{ar 3600 1010 58 58  0.0000000 6.2831853}%
% CIRCLE 2 0 0 0
% 4 2800 1820 2760 1860 2760 1860 2790 1830
% 
\special{pn 8}%
\special{sh 0.600}%
\special{ar 2800 1820 58 58  0.0000000 6.2831853}%
% CIRCLE 2 0 0 0
% 4 3610 2610 3570 2650 3570 2650 3600 2620
% 
\special{pn 8}%
\special{sh 0.600}%
\special{ar 3610 2610 58 58  0.0000000 6.2831853}%
% CIRCLE 2 0 0 0
% 4 5210 2620 5170 2660 5170 2660 5200 2630
% 
\special{pn 8}%
\special{sh 0.600}%
\special{ar 5210 2620 58 58  0.0000000 6.2831853}%
% CIRCLE 2 0 0 0
% 4 2010 2600 1970 2640 1970 2640 2000 2610
% 
\special{pn 8}%
\special{sh 0.600}%
\special{ar 2010 2600 58 58  0.0000000 6.2831853}%
% LINE 2 0 3 0
% 4 390 3180 6780 3190 6780 3190 6780 3190
% 
\special{pn 8}%
\special{pa 390 3180}%
\special{pa 6780 3190}%
\special{fp}%
\special{pa 6780 3190}%
\special{pa 6780 3190}%
\special{fp}%
% LINE 2 0 3 0
% 2 810 3790 800 320
% 
\special{pn 8}%
\special{pa 810 3790}%
\special{pa 800 320}%
\special{fp}%
% VECTOR 2 0 3 0
% 2 6490 3190 6830 3190
% 
\special{pn 8}%
\special{pa 6490 3190}%
\special{pa 6830 3190}%
\special{fp}%
\special{sh 1}%
\special{pa 6830 3190}%
\special{pa 6764 3170}%
\special{pa 6778 3190}%
\special{pa 6764 3210}%
\special{pa 6830 3190}%
\special{fp}%
% VECTOR 2 0 3 0
% 2 800 200 800 200
% 
\special{pn 8}%
\special{pa 800 200}%
\special{pa 800 200}%
\special{fp}%
% VECTOR 2 0 3 0
% 2 800 790 800 300
% 
\special{pn 8}%
\special{pa 800 790}%
\special{pa 800 300}%
\special{fp}%
\special{sh 1}%
\special{pa 800 300}%
\special{pa 780 368}%
\special{pa 800 354}%
\special{pa 820 368}%
\special{pa 800 300}%
\special{fp}%
% STR 2 0 3 0
% 3 3600 3300 3600 3400 5 0
% 0
\put(36.0000,-34.0000){\makebox(0,0){0}}%
% STR 2 0 3 0
% 3 4400 3280 4400 3380 5 0
% 1
\put(44.0000,-33.8000){\makebox(0,0){1}}%
% STR 2 0 3 0
% 3 5190 3290 5190 3390 5 0
% 2
\put(51.9000,-33.9000){\makebox(0,0){2}}%
% STR 2 0 3 0
% 3 2800 3300 2800 3400 5 0
% -1
\put(28.0000,-34.0000){\makebox(0,0){-1}}%
% STR 2 0 3 0
% 3 1990 3290 1990 3390 5 0
% -2
\put(19.9000,-33.9000){\makebox(0,0){-2}}%
% STR 2 0 3 0
% 3 600 2500 600 2600 5 0
% 1
\put(6.0000,-26.0000){\makebox(0,0){1}}%
% STR 2 0 3 0
% 3 600 1710 600 1810 5 0
% 3/2
\put(6.0000,-18.1000){\makebox(0,0){3/2}}%
% STR 2 0 3 0
% 3 600 890 600 990 5 0
% 2
\put(6.0000,-9.9000){\makebox(0,0){2}}%
% STR 2 0 3 0
% 3 6780 3270 6780 3370 5 0
% $q$
\put(67.8000,-33.7000){\makebox(0,0){$q$}}%
% STR 2 0 3 0
% 3 560 230 560 330 5 0
% $h$
\put(5.6000,-3.3000){\makebox(0,0){$h$}}%
% STR 2 0 3 0
% 3 3780 1080 3780 1180 2 0
% $T$
\put(37.8000,-11.8000){\makebox(0,0)[lb]{$T$}}%
% STR 2 0 3 0
% 3 3770 2680 3770 2780 5 0
% $J$
\put(37.7000,-27.8000){\makebox(0,0){$J$}}%
% STR 2 0 3 0
% 3 4720 1840 4720 1940 5 0
% $G^+$
\put(47.2000,-19.4000){\makebox(0,0){$G^+$}}%
% STR 2 0 3 0
% 3 3130 1840 3130 1940 5 0
% $G^-$
\put(31.3000,-19.4000){\makebox(0,0){$G^-$}}%
% STR 2 0 3 0
% 3 5430 2670 5430 2770 5 0
% $J^+$
\put(54.3000,-27.7000){\makebox(0,0){$J^+$}}%
% STR 2 0 3 0
% 3 4650 1510 4650 1610 5 0
% ${G'}^+$
\put(46.5000,-16.1000){\makebox(0,0){${G'}^+$}}%
% STR 2 0 3 0
% 3 2240 2650 2240 2750 5 0
% $J^-$
\put(22.4000,-27.5000){\makebox(0,0){$J^-$}}%
% STR 2 0 3 0
% 3 3030 1520 3030 1620 5 0
% ${G'}^-$
\put(30.3000,-16.2000){\makebox(0,0){${G'}^-$}}%
% CIRCLE 2 0 3 0
% 4 2800 1820 2810 1910 2810 1910 2810 1910
% 
\special{pn 8}%
\special{ar 2800 1820 92 92  0.0000000 6.2831853}%
% CIRCLE 2 0 0 0
% 4 4402 1812 4362 1852 4362 1852 4392 1822
% 
\special{pn 8}%
\special{sh 0.600}%
\special{ar 4402 1812 58 58  0.0000000 6.2831853}%
% CIRCLE 2 0 3 0
% 4 4401 1811 4411 1901 4411 1901 4411 1901
% 
\special{pn 8}%
\special{ar 4402 1812 92 92  0.0000000 6.2831853}%
\end{picture}%

%% file: paper.bbl
\begin{thebibliography}{99}
%\cite{Witten:2003nn}
\bibitem{Witten:2003nn}
  E.~Witten,
  ``Perturbative gauge theory as a string theory in twistor space,''
  Commun.\ Math.\ Phys.\  {\bf 252} (2004) 189
  [arXiv:hep-th/0312171].
  %%CITATION = HEP-TH 0312171;%%
%\cite{Gukov:2004ei}
\bibitem{Gukov:2004ei}
  S.~Gukov, L.~Motl and A.~Neitzke,
  ``Equivalence of twistor prescriptions for super Yang-Mills,''
  arXiv:hep-th/0404085.
  %%CITATION = HEP-TH 0404085;%%
%\cite{Popov:2004rb}
\bibitem{Popov:2004rb}
  A.~D.~Popov and C.~Saemann,
  ``On supertwistors, the Penrose-Ward transform and N = 4 super Yang-Mills
  theory,''
  Adv.\ Theor.\ Math.\ Phys.\  {\bf 9} (2005) 931
  [arXiv:hep-th/0405123].
  %%CITATION = HEP-TH 0405123;%%
%\cite{Lechtenfeld:2004cc}
\bibitem{Lechtenfeld:2004cc}
  O.~Lechtenfeld and A.~D.~Popov,
  ``Supertwistors and cubic string field theory for open N = 2 strings,''
  Phys.\ Lett.\ B {\bf 598} (2004) 113
  [arXiv:hep-th/0406179].
  %%CITATION = HEP-TH 0406179;%%
%\cite{Popov:2004nk}
\bibitem{Popov:2004nk}
  A.~D.~Popov and M.~Wolf,
  ``Topological B-model on weighted projective spaces and self-dual models  in
  four dimensions,''
  JHEP {\bf 0409} (2004) 007
  [arXiv:hep-th/0406224].
  %%CITATION = HEP-TH 0406224;%%
%\cite{Siegel:2004dj}
\bibitem{Siegel:2004dj}
  W.~Siegel,
  ``Untwisting the twistor superstring,''
  arXiv:hep-th/0404255.
  %%CITATION = HEP-TH 0404255;%%
%\cite{Sinkovics:2004fm}
\bibitem{Sinkovics:2004fm}
  A.~Sinkovics and E.~P.~Verlinde,
  ``A six dimensional view on twistors,''
  Phys.\ Lett.\ B {\bf 608} (2005) 142
  [arXiv:hep-th/0410014].
  %%CITATION = HEP-TH 0410014;%%
%\cite{Movshev:2004ub}
\bibitem{Movshev:2004ub}
  M.~Movshev,
  ``Yang-Mills theory and a superquadric,''
  arXiv:hep-th/0411111.
  %%CITATION = HEP-TH 0411111;%%
%\cite{Wolf:2004hp}
\bibitem{Wolf:2004hp}
  M.~Wolf,
  ``On hidden symmetries of a super gauge theory and twistor string theory,''
  JHEP {\bf 0502} (2005) 018
  [arXiv:hep-th/0412163].
  %%CITATION = HEP-TH 0412163;%%
%\cite{Seki:2005hx}
\bibitem{Seki:2005hx}
  S.~Seki and K.~Sugiyama,
  ``Gauged linear sigma model on supermanifold,''
  arXiv:hep-th/0503074.
  %%CITATION = HEP-TH 0503074;%%
%\cite{Tokunaga:2005pj}
\bibitem{Tokunaga:2005pj}
  T.~Tokunaga,
  ``String theories on flat supermanifolds,''
  arXiv:hep-th/0509198.
  %%CITATION = HEP-TH 0509198;%%
%\cite{Sethi:1994ch}
\bibitem{Sethi:1994ch}
  S.~Sethi,
  ``Supermanifolds, rigid manifolds and mirror symmetry,''
  Nucl.\ Phys.\ B {\bf 430} (1994) 31
  [arXiv:hep-th/9404186].
  %%CITATION = HEP-TH 9404186;%%
%\cite{Schwarz:1995ak}
\bibitem{Schwarz:1995ak}
  A.~S.~Schwarz,
  ``Sigma models having supermanifolds as target spaces,''
  Lett.\ Math.\ Phys.\  {\bf 38} (1996) 91
  [arXiv:hep-th/9506070].
  %%CITATION = HEP-TH 9506070;%%
%\cite{Aganagic:2004yh}
\bibitem{Aganagic:2004yh}
  M.~Aganagic and C.~Vafa,
  ``Mirror symmetry and supermanifolds,''
  arXiv:hep-th/0403192.
  %%CITATION = HEP-TH 0403192;%%
%\cite{Kumar:2004dj}
\bibitem{Kumar:2004dj}
  S.~P.~Kumar and G.~Policastro,
  ``Strings in twistor superspace and mirror symmetry,''
  Phys.\ Lett.\ B {\bf 619} (2005) 163
  [arXiv:hep-th/0405236].
  %%CITATION = HEP-TH 0405236;%%
%\cite{Ahn:2004xs}
\bibitem{Ahn:2004xs}
  C.~h.~Ahn,
  ``Mirror symmetry of Calabi-Yau supermanifolds,''
  Mod.\ Phys.\ Lett.\ A {\bf 20} (2005) 407
  [arXiv:hep-th/0407009].
  %%CITATION = HEP-TH 0407009;%%
%\cite{Belhaj:2004ts}
\bibitem{Belhaj:2004ts}
  A.~Belhaj, L.~B.~Drissi, J.~Rasmussen, E.~H.~Saidi and A.~Sebbar,
  ``Toric Calabi-Yau supermanifolds and mirror symmetry,''
  J.\ Phys.\ A {\bf 38} (2005) 6405
  [arXiv:hep-th/0410291].
  %%CITATION = HEP-TH 0410291;%%
%\cite{AhlLaamara:2006rk}
\bibitem{AhlLaamara:2006rk}
  R.~Ahl Laamara, A.~Belhaj, L.~B.~Drissi and E.~H.~Saidi,
  ``On local Calabi-Yau supermanifolds and their mirrors,''
  arXiv:hep-th/0601215.
  %%CITATION = HEP-TH 0601215;%%
%\cite{Kaura:2006hb}
\bibitem{Kaura:2006hb}
  P.~Kaura, A.~Misra and P.~Shukla,
  ``Super Picard-Fuchs equation and monodromies for supermanifolds,''
  arXiv:hep-th/0603126.
  %%CITATION = HEP-TH 0603126;%%
%\cite{Berkovits:2004hg}
\bibitem{Berkovits:2004hg}
  N.~Berkovits,
  ``An alternative string theory in twistor space for N = 4
  super-Yang-Mills,''
  Phys.\ Rev.\ Lett.\  {\bf 93} (2004) 011601
  [arXiv:hep-th/0402045].
  %%CITATION = HEP-TH 0402045;%%
%\cite{Berkovits:2004tx}
\bibitem{Berkovits:2004tx}
  N.~Berkovits and L.~Motl,
  ``Cubic twistorial string field theory,''
  JHEP {\bf 0404} (2004) 056
  [arXiv:hep-th/0403187].
  %%CITATION = HEP-TH 0403187;%%
%\cite{Berkovits:2004jj}
\bibitem{Berkovits:2004jj}
  N.~Berkovits and E.~Witten,
  ``Conformal supergravity in twistor-string theory,''
  JHEP {\bf 0408} (2004) 009
  [arXiv:hep-th/0406051].
  %%CITATION = HEP-TH 0406051;%%
%\cite{Ahn:2004yu}
\bibitem{Ahn:2004yu}
  C.~h.~Ahn,
  ``N = 1 conformal supergravity and twistor-string theory,''
  JHEP {\bf 0410} (2004) 064
  [arXiv:hep-th/0409195].
  %%CITATION = HEP-TH 0409195;%%
%\cite{Park:2004bw}
\bibitem{Park:2004bw}
  J.~Park and S.~J.~Rey,
  ``Supertwistor orbifolds: Gauge theory amplitudes and topological  strings,''
  JHEP {\bf 0412} (2004) 017
  [arXiv:hep-th/0411123].
  %%CITATION = HEP-TH 0411123;%%
%\cite{Giombi:2004xv}
\bibitem{Giombi:2004xv}
  S.~Giombi, M.~Kulaxizi, R.~Ricci, D.~Robles-Llana, D.~Trancanelli and K.~Zoubos,
  ``Orbifolding the twistor string,''
  Nucl.\ Phys.\ B {\bf 719} (2005) 234
  [arXiv:hep-th/0411171].
  %%CITATION = HEP-TH 0411171;%%
%\cite{Ahn:2004ua}
\bibitem{Ahn:2004ua}
  C.~h.~Ahn,
  ``N = 2 conformal supergravity from twistor-string theory,''
  arXiv:hep-th/0412202.
  %%CITATION = HEP-TH 0412202;%%
%\cite{Grassi:2004tv}
\bibitem{Grassi:2004tv}
  P.~A.~Grassi and G.~Policastro,
  ``Super-Chern-Simons theory as superstring theory,''
  arXiv:hep-th/0412272.
  %%CITATION = HEP-TH 0412272;%%
%\cite{Saemann:2005ji}
\bibitem{Saemann:2005ji}
  C.~Saemann,
  ``On the mini-superambitwistor space and N = 8 super Yang-Mills theory,''
  arXiv:hep-th/0508137.
  %%CITATION = HEP-TH 0508137;%%
%\cite{Lechtenfeld:2005xi}
\bibitem{Lechtenfeld:2005xi}
  O.~Lechtenfeld and C.~Saemann,
  ``Matrix models and D-branes in twistor string theory,''
  JHEP {\bf 0603} (2006) 002
  [arXiv:hep-th/0511130].
  %%CITATION = HEP-TH 0511130;%%
%\cite{Neitzke:2004pf}
\bibitem{Neitzke:2004pf}
  A.~Neitzke and C.~Vafa,
  ``N = 2 strings and the twistorial Calabi-Yau,''
  arXiv:hep-th/0402128.
  %%CITATION = HEP-TH 0402128;%%
%\cite{Nekrasov:2004js}
\bibitem{Nekrasov:2004js}
  N.~Nekrasov, H.~Ooguri and C.~Vafa,
  ``S-duality and topological strings,''
  JHEP {\bf 0410} (2004) 009
  [arXiv:hep-th/0403167].
  %%CITATION = HEP-TH 0403167;%%
%\cite{Rocek:2004bi}
\bibitem{Rocek:2004bi}
  M.~Rocek and N.~Wadhwa,
  ``On Calabi-Yau supermanifolds,''
  arXiv:hep-th/0408188.
  %%CITATION = HEP-TH 0408188;%%
%\cite{Zhou:2004su}
\bibitem{Zhou:2004su}
  C.~g.~Zhou,
  ``On Ricci flat supermanifolds,''
  JHEP {\bf 0502} (2005) 004
  [arXiv:hep-th/0410047].
  %%CITATION = HEP-TH 0410047;%%
%\cite{Witten:1993yc}
\bibitem{Witten:1993yc}
  E.~Witten,
  ``Phases of N = 2 theories in two dimensions,''
  Nucl.\ Phys.\ B {\bf 403} (1993) 159
  [arXiv:hep-th/9301042].
  %%CITATION = HEP-TH 9301042;%%
%\cite{Konishi:1983hf}
\bibitem{Konishi:1983hf}
  K.~Konishi,
  ``Anomalous Supersymmetry Transformation Of Some Composite Operators In
  Sqcd,''
  Phys.\ Lett.\ B {\bf 135} (1984) 439.
  %%CITATION = PHLTA,B135,439;%%
%\cite{Konishi:1985tu}
\bibitem{Konishi:1985tu}
  K.~i.~Konishi and K.~i.~Shizuya,
  ``Functional Integral Approach To Chiral Anomalies In Supersymmetric Gauge
  Theories,''
  Nuovo Cim.\ A {\bf 90} (1985) 111.
  %%CITATION = NUCIA,A90,111;%%
%\cite{Basu:2003bq}
\bibitem{Basu:2003bq}
  A.~Basu and S.~Sethi,
  ``World-sheet stability of (0,2) linear sigma models,''
  Phys.\ Rev.\ D {\bf 68} (2003) 025003
  [arXiv:hep-th/0303066].
  %%CITATION = HEP-TH 0303066;%%
%\cite{Hori:2001ax}
\bibitem{Hori:2001ax}
  K.~Hori and A.~Kapustin,
  ``Duality of the fermionic 2d black hole and N = 2 Liouville theory as
  mirror symmetry,''
  JHEP {\bf 0108} (2001) 045
  [arXiv:hep-th/0104202].
  %%CITATION = HEP-TH 0104202;%%
\end{thebibliography}
